\documentclass[aip,amsmath,amssymb,preprint,jcp]{revtex4-1} 

\usepackage{graphicx}
\usepackage{dcolumn}
\usepackage{bm}

\usepackage[utf8]{inputenc}
\usepackage[T1]{fontenc}
\usepackage{mathptmx}
\usepackage{etoolbox}

\usepackage{derivative}
\usepackage[mathscr]{eucal}
\usepackage{interval}
\usepackage{mathtools}
\usepackage{physics2}

\allowdisplaybreaks{}
\usephysicsmodule{ab,ab.legacy,braket,op.legacy,qtext.legacy}
\DeclarePairedDelimiter{\floor}{\lfloor}{\rfloor}
\newcommand{\sch}{Schr\"{o}dinger}
\newcommand{\e}[1]{\mathrm{e}^{#1}}
\newcommand{\iu}{\mathrm{i}}
\newcommand{\ad}{\mathrm{a}}
\newcommand{\dia}{\mathrm{d}}
\newcommand{\ev}[1]{\braket[1]{#1}}
\newcommand{\LP}[1]{\mathcal{L}^{P}_{\text{#1}}}
\newcommand{\LPM}[1]{\mathcal{M}_{\text{#1}}}
\newcommand{\tint}[1]{\int_0^{\tilde{t}}#1\odif{\tilde{\tau}}}

\makeatletter
\def\@email#1#2{%
	\endgroup
	\patchcmd{\titleblock@produce}
		{\frontmatter@RRAPformat}
		{\frontmatter@RRAPformat{\produce@RRAP{*#1\href{mailto:#2}{#2}}}\frontmatter@RRAPformat}
		{}{}
}%
\makeatother

\begin{document}

	\title[Negative Marginal Densities in Mixed Quantum\textendash{}Classical Liouville Dynamics]{Negative Marginal Densities in Mixed Quantum\textendash{}Classical Liouville Dynamics}
	\author{Kai Gu}
	\author{Jeremy Schofield}
		\affiliation{Chemical Physics Theory Group, Department of Chemistry, University of Toronto, Toronto, Ontario, Canada M5S 3H6}
		\email{jeremy.schofield@utoronto.ca}
	\date{\today}

	\begin{abstract}
		The mixed quantum\textendash{}classical Liouville equation (QCLE) provides an approximate perturbative framework for describing the dynamics of systems with coupled quantum and classical degrees of freedom of disparate thermal wavelengths. The evolution governed by the Liouville operator preserves many properties of full quantum dynamics, including the conservation of total population, energy, and purity, and has shown quantitative agreement with exact quantum results for the expectation values of many observables where direct comparisons are feasible. However, since the QCLE density matrix operator is obtained from the partial Wigner transform of the full quantum density matrix, its matrix elements can have negative values, implying that the diagonal matrix elements behave as pseudo-densities rather than densities of classical phase space. Here, we compare phase-space distributions generated by exact quantum dynamics with those produced by QCLE evolution from pure quantum initial states. We show that resonance effects in the off-diagonal matrix elements differ qualitatively, particularly for low-energy states. Furthermore, numerical and analytical results for low-dimensional models reveal that the QCLE can violate the positivity of marginal phase-space densities, a property that should hold at all times for any physical system. A perturbative analysis of a model system confirms that such violations arise generically. We also show that the violations of positivity of the marginal densities vanish as the initial energy of the system increases relative to the energy gap between subsystem states. These findings suggest that a negativity index, quantifying deviations from positivity, may provide a useful metric for assessing the validity of mixed quantum\textendash{}classical descriptions.
	\end{abstract}


	\maketitle

	\section{Introduction}
		Processes such as electron and proton transfer\cite{bell73, doi:10.1142/3347, Langford2025}, excited-state relaxation\cite{femtochem01, DARUGAR2003284, C7CP08696B}, and energy transport in light-harvesting complexes\cite{annurev:/content/journals/10.1146/annurev.physchem.040808.090259, doi:10.1021/jz900062f, Ullah2022, Wu2025} are intrinsically quantum mechanical, making the development of theoretical and computational approaches for quantum many-body systems a central research focus. Although fully quantum approaches offer the most accurate description of the dynamics of large, complex systems, their application is often impractical because the computational cost of solving the \sch{} equation scales exponentially with system size.

		The mixed quantum\textendash{}classical Liouville equation (QCLE)\cite{Aleksandrov1981, Gerasimenko1982, FILINOV200819961, FILINOV200819962, 10.1063/1.478811, annurev:/content/journals/10.1146/annurev.physchem.57.032905.104702} emerges as a systematic alternative, derived rigorously from the partial Wigner representation of the quantum Liouville equation through a controlled expansion in a small parameter defined by the ratio of thermal de Broglie wavelengths of the environment and subsystem. Unlike phenomenological dynamical models, the QCLE constitutes a principled approximation to the exact quantum dynamics, offering a direct and transparent connection between quantum-mechanical and hybrid quantum\textendash{}classical descriptions. Importantly, it incorporates the back-reaction of the quantum subsystem on the classical environment, preserves the total energy of the system, and maintains invariance under canonical transformations, ensuring that the dynamics are independent of the particular choice of canonical coordinates used to represent the classical surroundings. This property underscores the theoretical robustness of the approach. In the appropriate limits, the QCLE reduces to the classical Liouville equation, yet it still captures essential quantum features such as coherence and nonadiabatic effects in the quantum subsystem. In this way, the QCLE occupies a well-defined position in the hierarchy of dynamical methods, bridging the gap between quantum and classical descriptions while preserving key aspects of quantum behavior in a computationally tractable form.

		Despite its theoretical strengths, direct numerical solution of the QCLE remains challenging due to the high dimensionality of the classical phase space and the need to resolve quantum\textendash{}classical correlations\cite{https://doi.org/10.1002/jcc.26045}. To overcome these obstacles, various approximate methods have been developed, including trajectory-based schemes\cite{10.1063/1.1313525, 10.1063/1.1425835} such as classical Liouville dynamics with quantum transitions and surface hopping\cite{10.1063/1.1433502, annurev:/content/journals/10.1146/annurev-physchem-040215-112245}, as well as phase-space mapping approaches\cite{doi:https://doi.org/10.1002/0471739464.ch5, 10.1063/1.2971041, doi:10.1139/V09-041, 10.1063/1.3480018, 10.1063/1.5143412, annurev:/content/journals/10.1146/annurev-physchem-082423-120631}. These methods balance accuracy and efficiency, enabling studies of energy transfer\cite{doi:10.1021/jz900062f, doi:10.1021/ar400263p, doi:10.1021/acs.jctc.5c01002}, reaction dynamics\cite{10.1063/1.1566731, 10.1063/1.4857335}, and spectroscopy in large systems\cite{doi:10.1073/pnas.0408813102, doi:10.1021/acs.jpca.5c02171}. Recent advances in machine learning have further expanded the applicability of QCLE-based simulations\cite{10.1063/5.0073689}.

		However, the QCLE and its approximations suffer from several inherent limitations. Trajectory-based methods, though efficient, often rely on ad hoc corrections to enforce detailed balance\cite{doi:10.1142/97898128121620002, doi:10.1142/97898128396640021} or to accurately represent quantum back-reaction\cite{10.1063/1.478811, doi:10.1021/jp0761416}. The classical treatment of the environment neglects quantum decoherence and zero-point energy effects, which can lead to unphysical energy flow and inaccurate long-time dynamics. Besides, the QCLE implicitly presupposes a clear separation between quantum and classical subsystems\textemdash{}an assumption that may fail in strongly correlated systems where quantum entanglement pervades the entire system.

		Historically, validation of the QCLE has relied mainly on reduced observables such as population transfer\cite{10.1063/1.1313525}, rates of coherence decay\cite{10.1063/1.1597496}, and reaction rates\cite{0chap-kapral05}, which are compared with exact quantum results or experimental data. While informative, such comparisons do not examine the full classical phase-space structure of the quantum\textendash{}classical system. Systematic comparisons of QCLE-derived marginal densities with those obtained from the time-dependent \sch{} equation have not been reported, and most studies typically lack the granularity needed to assess the true accuracy of the QCLE, particularly with respect to the coherence structure. This omission is important because agreement in reduced properties does not guarantee physical consistency.

		Moreover, previous studies have rarely delineated the boundaries of QCLE applicability, especially in regimes ominated by quantum bath effects, such as low temperatures, strong tunneling, or cases where back-reaction induces nonclassical distortions in the environment. The absence of phase-space-based validation and of well-defined applicability criteria leaves unresolved questions regarding the self-consistency and interpretive scope of the QCLE framework.

		In many applications, one is interested in the statistics of a reduced set of observable quantities associated with experimental measurements. In such situations, a common approach is to eliminate irrelevant degrees of freedom using a projection-operator approach to obtain the generalized quantum master equation, also known as the Nakajima\textendash{}Zwanzig equation\cite{10.1143/PTP.20.948, 10.1063/1.1731409}. Since this equation is typically as difficult to solve as the original quantum problem, many practical implementations\textemdash{}including the Redfield\cite{REDFIELD19651} and Lindblad equations\cite{10.1093/acprof:oso/9780199213900.001.0001}\textemdash{}are based on approximate expansions of the time-convolutionless (TCL) operator or on ad hoc Markovian closures that may fail to preserve complete positivity of the density matrix, leading to unphysical results such as negative populations\cite{10.1063/1.463831}. These problems often stem from inconsistent truncation schemes or the use of approximations without enforcing restrictions that ensure positivity. While such methods can provide useful short-time behavior at reduced computational cost, their tendency to violate positivity at long times highlights a fundamental trade-off between rigor and efficiency. In some cases, such as the Redfield Markovian equation, positivity of the determinant of the density matrix can be restored with slippage superoperators that construct physically-allowable initial conditions of the reduced density matrix\cite{10.1063/1.463831, 10.1063/1.479867}.

		Against this backdrop of trade-offs and known pathologies, we identify an unreported defect in the QCLE:~the loss of positivity not in the full density matrix, but in its marginal classical phase-space densities. In Sec.~\ref{sec:method}, we review the theoretical background of nonadiabatic dynamics in both configuration and phase space, introducing the QCLE and the concept of marginal densities. Sec.~\ref{sec:DAC} presents numerical results showing that, although reduced observables such as populations computed using the QCLE and the time-dependent \sch{} equation agree, their corresponding phase-space densities differ significantly, leading to nonphysical discrepancies in the marginals. In addition to these numerical results, Sec.~\ref{sec:const model} provides an analytical model demonstrating the breakdown of positive semidefiniteness in the marginal density. In Sec.~\ref{sec:conc}, the origin and the significance of this breakdown are discussed.

	\section{\label{sec:method}Mixed Quantum\textendash{}Classical Systems}
		In this section, we establish notation and provide a brief overview of the partial Wigner transform, which serves as the foundation for deriving the QCLE.~The QCLE emerges from applying the partial Wigner transform to the Liouville\textendash{}von Neumann equation for the density matrix\cite{Kapral_2015}. This procedure preserves the operator nature of the quantum subsystem, typically expressed in either the diabatic or adiabatic basis, as discussed below.

		\subsection{Diabatic and Adiabatic States}
			We consider a system composed of a heavy bath particle of mass \(M\), with position and momentum operators \(\hat{R}\) and \(\hat{P}\), respectively, and a subsystem of lighter particles described by an isolated Hamiltonian operator \(\hat{H}_s\). The system interacts with the bath degrees of freedom through a potential \(\hat{V}_c(\hat{r},\hat{R})\). The total Hamiltonian can be written as
			\begin{equation}
				\hat{H}=\hat{T}_R+\hat{H}_0(\hat{r};\hat{R}) ,
			\end{equation}
			where \(\hat{T}_R=\hat{P}^2/(2M)=-\hbar^2/(2M)\pdv[2]{}/{R}\) is the bath kinetic energy operator and \(\hat{H}_0=\hat{H}_s+\hat{V}_c(\hat{r},\hat{R})\) contains the remaining terms of the total Hamiltonian, generally including the bath potential, the subsystem kinetic and potential energy, and the bath-subsystem interaction \(\hat{V}_c(\hat{r},\hat{R})\).

			The eigenstates of \(\hat{H}_0\) define the adiabatic basis,
			\begin{equation}
				\hat{H}_0\ket{\alpha (R)}=E_{\alpha}(R)\ket{\alpha (R)} ,
			\end{equation}
			while under the \(R\)-independent diabatic basis \(\ket{i}\), the diabatic potential matrix \(V (R)\) is
			\begin{equation}
				V_{ij}(R)=\braket[3]{i}{\hat{H}_0(r,R)}{j} ,
			\end{equation}
			and the total wavefunction can be expanded in terms of either the diabatic or the adiabatic basis,
			\begin{subequations}
				\begin{align}
					\ket{\Psi(R,t)}&=\sum_{\alpha}\psi_{\alpha}(R,t)\ket{\alpha (R)} ,\\
					&=\sum_{i}\psi_i(R,t)\ket{i} .
				\end{align}
			\end{subequations}
			In what follows, a matrix element of an operator represented in the adiabatic basis will be indicated with Greek letter indices, while Roman indices will indicate diabatic matrix elements.

			The transformation matrix \(U\) that converts a state vector from the adiabatic basis into the diabatic basis, provided \(R\)-independent basis states exist\cite{PhysRev.179.111, Baer01071980, 10.1063/1.443853}, is defined as
			\begin{equation}
				U_{i\alpha}(R)=\braket{i}{\alpha (R)} .
			\end{equation}
			For simplicity, we focus on a two-level system in which only the ground and first excited state participate in the dynamics. For such a two-level system, the matrix \(U\) exists and is given by\cite{10.1063/1.443853}
			\begin{equation}
				U(R)=\begin{pmatrix}\cos(\theta(R)/2)&-\sin(\theta(R)/2)\\\sin(\theta(R)/2)&\cos(\theta(R)/2)\end{pmatrix} ,
			\end{equation}
			where the mixing angle \(\theta\) is associated with the diabatic Hamiltonian via
			\begin{equation}
				\theta(R)=\arctan\ab(\frac{2V_{01}(R)}{V_{00}(R)-V_{11}(R)}) .
			\end{equation}

			Unless specified, we limit discussion to the case of a one-dimensional bath. Extension to multiple degrees of freedom and higher spatial dimensions is straightforward.

		\subsection{Mixed Quantum\textendash{}Classical Dynamics}
			The partial Wigner transform of the density matrix is defined as\cite{10.1063/1.478811}
			\begin{equation}
				\hat{\rho}_W(R,P,t)=\frac{1}{\pi\hbar}\int\braket[3]{R-Q}{\hat{\rho}(t)}{R+Q}\e{2\iu PQ/\hbar}\odif{Q} ,
			\end{equation}
			which maps the quantum density matrix to a subsystem density operator with classical phase space coordinates \(R\) and \(P\) for a subset of degrees of freedom chosen by their classical characteristics due to their relatively short thermal wavelengths. Similarly, quantum operators \(\hat{B}\) are transformed as,
			\begin{equation}
				\hat{B}_W(R,P)=2\int\braket[3]{R-Q}{\hat{B}}{R+Q}\e{2\iu PQ/\hbar}\odif{Q} ,
			\end{equation}
			which yield operators in the quantum subsystem. In the equations above, the integral over the coordinate \(Q\) extends over the physically allowed positions, which are taken to extend over the entire real axis, and this convention applies hereafter unless specified otherwise.

			The product of two operators is transformed to\cite{Groenewold1946, 10.1063/1.1705323}
			\begin{equation}
				{(\hat{A}\hat{B})}_W(R,P)=\hat{A}_W(R,P)\e{\hbar\Lambda/(2\iu)}\hat{B}_W(R,P) ,
			\end{equation}
			where
			\begin{equation}
				\Lambda=\overleftarrow{\pdv{}{P}}\overrightarrow{\pdv{}{R}}-\overleftarrow{\pdv{}{R}}\overrightarrow{\pdv{}{P}}
			\end{equation}
			is the negative of the Poisson bracket operator, and the arrows indicate the direction in which the derivatives act.

			The evolution of the quantum\textendash{}classical density matrix \(\hat{\rho}_W\) can be derived by taking the partial Wigner transform of the Liouville\textendash{}von Neumann equation to obtain a formally exact expression for the evolution of \(\hat{\rho}_W\)\cite{10.1063/1.478811},
			\begin{equation}\label{eq:QLE}
				\pdv{\hat{\rho}_W}{t}=-\frac{\iu}{\hbar}\ab(\hat{H}_W\e{\hbar\Lambda/(2\iu)}\hat{\rho}_W-\hat{\rho}_W\e{\hbar\Lambda/(2\iu)}\hat{H}_W) .
			\end{equation}
			Noting that the small mass ratio \(m/M\) of the subsystem and bath particles implies that the thermal wavelength of the bath particles \(\lambda_M=\hbar/\sqrt{M k_\text{B}T}\) is much shorter than that of the subsystem particles \(\lambda_m=\hbar/\sqrt{m k_\text{B}T}\), Eq.~\ref{eq:QLE} can be expressed in scaled coordinates and perturbatively expanded in this mass ratio to obtain the mixed quantum\textendash{}classical Liouville equation (QCLE)\cite{10.1063/1.478811, 10.1063/1.1705323},
			\begin{align}\label{eq:QCLE}
				\pdv{\hat{\rho}_W}{t}&=-\frac{\iu}{\hbar}[\hat{H}_W,\hat{\rho}_W]+\frac{1}{2}\ab(\{\hat{H}_W,\hat{\rho}_W\}-\{\hat{\rho}_W,\hat{H}_W\})\nonumber\\
				&=-\iu\hat{\mathcal{L}}\hat{\rho}_W,
			\end{align}
			which defines the mixed quantum\textendash{}classical Liouville operator \(\iu\hat{\mathcal{L}}\). In the scaled coordinates, the corrections to Eq.~\ref{eq:QCLE} are \(\mathcal{O}(m/M)\). Note that this equation retains the Hilbert space character of the subsystem, which may be represented in any complete basis for this space.

			In the adiabatic basis, elements of the partial Wigner-transformed density matrix (PWTDM)
			\begin{equation}
				\rho_{W}^{\alpha\beta}(R,P,t)=\braket[3]{\alpha(R)}{\hat{\rho}_W(R,P,t)}{\beta(R)}
			\end{equation}
			evolve by
			\begin{multline}\label{eq:adia QCLE}
				\pdv{\rho_{W}^{\alpha\beta}}{t}=-\iu\omega_{\alpha\beta}\rho_{W}^{\alpha\beta}+\sum_{\gamma}\frac{P}{M}(\rho_{W}^{\alpha\gamma}d_{\gamma\beta}-d_{\alpha\gamma}\rho_W^{\gamma\beta})\\
				-\frac{P}{M}\pdv{\rho_{W}^{\alpha\beta}}{R}-\frac{1}{2}\sum_{\gamma}\ab(F_W^{\alpha\gamma}\pdv{\rho_{W}^{\gamma\beta}}{P}+\pdv{\rho_W^{\alpha\gamma}}{P}F_W^{\gamma\beta}) ,
			\end{multline}
			where \(\omega_{\alpha\beta}=(E_{\alpha}-E_{\beta})/\hbar\), \(d_{\alpha\beta}(R)\) is the first-order nonadiabatic coupling matrix element, and \(F^{\alpha\beta}_W(R)\) is the adiabatic force matrix, defined as
			\begin{align}
				d_{\alpha\beta}(R)&=\braket[3]{\alpha(R)}{\pdv{}{R}}{\beta(R)} ,\\
				F^{\alpha\beta}_W(R)&=-\braket[3]{\alpha(R)}{\pdv{V}{R}}{\beta(R)} .
			\end{align}
			The diagonal element of the force matrix, \(F^{\alpha\alpha}_W(R)=-\pdv{E_\alpha}/{R}\), is the Hellmann-Feynman force for the adiabatic state \(\alpha\). Note that the density matrix is Hermitian and obeys \(\rho_W^{\beta\alpha}={\big(\rho_W^{\alpha\beta}\big)}^*\) at all times.

			The connection between the quantum density matrix and a representation of the mixed quantum\textendash{}classical density matrix operator for a pure state differs in the diabatic and adiabatic bases. If the wavefunction is written in the diabatic representation, the PWTDM is the transform of a product of diabatic wavefunctions on different surfaces,
			\begin{equation}\label{eq:pwtdm diabatic}
				\rho_{W}^{ij}(R,P,t)=\frac{1}{\pi\hbar}\int\psi_i(R-Q,t)\psi^*_j(R+Q,t)\e{2\iu PQ/\hbar}\odif{Q} .
			\end{equation}
			However, the construction of the density matrix from an adiabatic wavefunction is complicated by the fact that the adiabatic basis is parametrically dependent on the bath coordinate \(R\). This coordinate dependence introduces a number of complications, such as geometric phases\cite{10.1063/1.4866366}. Here, it follows that the matrix elements of the PWTDM in the adiabatic basis are \textit{not} the partial Wigner transform of a product of adiabatic wavefunctions. Instead, the adiabatic wavefunction must first be expanded in the \(R\)-independent diabatic basis to enable the construction of the partial Wigner transform of the density matrix operator there. The adiabatic representation of the transformed diabatic density matrix is then obtained by applying the unitary transformation converting the diabatic basis to the adiabatic basis:
			\begin{subequations}\label{eqs:WT}
				\begin{align}
					\psi_i(R,t)&=\sum_{\alpha}U_{i\alpha}(R)\psi_{\alpha}(R,t) ,\\
					\rho_{W}^{\alpha\beta}(R,P,t)&=\sum_{i,j}U^*_{i\alpha}(R)\rho_{W}^{ij}(R,P,t)U_{j\beta}(R) ,
				\end{align}
			\end{subequations}
			where \(\rho_{W}^{ij}\) is the PWTDM in diabatic basis defined in Eq.~\ref{eq:pwtdm diabatic}.

			In a mixed quantum\textendash{}classical system, the average of an observable \(\hat{O}_W\) is basis-independent:
			\begin{subequations}
				\begin{align}
					\ev{\hat{O}_W}&=\Tr\int\hat{O}_W(R,P)\hat{\rho}_W(R,P,t)\odif{R,P}\\
					&=\sum_{\alpha,\beta}\int O_{W}^{\alpha\beta}(R,P)\rho_{W}^{\beta\alpha}(R,P,t)\odif{R,P}\\
					&=\sum_{i,j}\int O_{W}^{ij}(R,P)\rho_{W}^{ji}(R,P,t)\odif{R,P} .
				\end{align}
			\end{subequations}

			Although approximate, the QCLE respects most of the conservation principles that follow from the \sch{} equation, as discussed in detail in Appendix~\ref{app:QCLE conserve}, including the conservation of population
			\begin{subequations}
			\begin{equation}
				\pdv{\ev{1}}{t}=\pdv{}{t}\Tr\int\hat{\rho}_W(R,P,t)\odif{R,P}=0 ,
			\end{equation}
			energy
			\begin{equation}
				\pdv{\ev{E}}{t}=\pdv{}{t}\Tr\int\hat{H}_W(R,P)\hat{\rho}_W(R,P,t)\odif{R,P}=0 ,
			\end{equation}
			and purity
			\begin{equation}
				\pdv{S}{t}=0 ,
			\end{equation}
			\end{subequations}
			where the purity \(S\) is defined as
			\begin{equation}\label{eq:def purity}
				S=\Tr\hat{\rho}^2=2\pi\hbar\Tr\int\hat{\rho}_W^2(R,P,t)\odif{R,P} .
			\end{equation}

			As its name suggests, the QCLE is a mixture of classical and quantum levels of description, aiming to treat the bath as a fully classical system, where position and momentum can be simultaneously measured, and the phase space density is positive-definite in all phase space. Since the QCLE is an approximation of the true quantum dynamics, it does not satisfy all the physical requirements that follow from the \sch{} equation, and important violations arise in physically observable properties of the bath degrees of freedom. Several deficiencies of the QCLE have been previously noted, particularly in the statistical mechanics of mixed quantum\textendash{}classical systems\cite{10.1063/1.1400129}. In the following, we show that for some systems of low dimensionality, the evolution under the QCLE leads to unphysical behavior that appears in observable quantities, such as negative ``probabilities'' for certain values of the position and/or the momentum of a bath degree of freedom, even if the initial state is a pure quantum state.

			To demonstrate that unphysical characteristics of the system may arise in QCLE dynamics, and to better understand the limits of the validity of this evolution equation, we consider the statistical properties of the bath degrees of freedom. The phase-space pseudo-density \(\rho(R,P,t)\) of the bath degrees of freedom in a mixed quantum\textendash{}classical system is computed by taking the trace of the PWTDM,
			\begin{equation}
				\rho(R,P,t)=\Tr\hat{\rho}_W(R,P,t)=\sum_\alpha\rho_W^{\alpha\alpha}(R,P,t)=\sum_i\rho_W^{ii}(R,P,t) ,
			\end{equation}
			and the corresponding (overall) marginal pseudo-densities of the bath position and momentum are
			\begin{subequations}
				\begin{align}
					n(R,t)&=\sum_{\alpha}\int\rho_W^{\alpha\alpha}(R,P,t)\odif{P}=\sum_{i}\int\rho_W^{ii}(R,P,t)\odif{P} ,\\
					\eta(P,t)&=\sum_{\alpha}\int\rho_W^{\alpha\alpha}(R,P,t)\odif{R}=\sum_{i}\int\rho_W^{ii}(R,P,t)\odif{R} ,
				\end{align}
			\end{subequations}
			which are independent of representation. The marginal densities are the probability to observe a bath position around \(R\) or a momentum value around \(P\), while the phase-space pseudo-density represents their joint probability. It has long been realized that the pseudo-density is not a true density since it is not positive-definite and is negative in regions of non-zero measure in phase space\cite{PhysRev.40.749}. If the bath were truly classical, then the position and momentum could be simultaneously measured with arbitrary precision, and the phase space density would be positive semidefinite. Thus, the emergence of negativities is an indicator of quantum effects in the bath. However, one might expect that the marginal densities \(n(R,t)\) and \(\eta(P,t)\) are positive-definite because integration of the trace of the partially-transformed density matrix relates the marginal densities to the corresponding positive quantum probabilities,
			\begin{subequations}\label{eqs:marginals}
				\begin{align}
					n(R,t)&=|\Psi(R,t)|^2=\sum_{i}|\psi_{i}(R,t)|^2 ,\\
					\eta(P,t)&=|\Phi(P,t)|^2=\sum_{i}\abs{\frac{1}{\sqrt{2\pi\hbar}}\int\psi_{i}(R,t)\e{-\iu PR/\hbar}\odif{R}}^2 .
				\end{align}
			\end{subequations}

			We may also define the marginal of a surface in the adiabatic basis (or, analogously, the diabatic basis), as well as a ``marginal'' of the off-diagonal elements, known as the coherence, as
			\begin{subequations}
				\begin{align}
					n_{\alpha}(R,t)&=\int\rho_W^{\alpha\alpha}(R,P,t)\odif{P} ,\\
					\eta_{\alpha}(P,t)&=\int\rho_W^{\alpha\alpha}(R,P,t)\odif{R} ,\\
					n_{\alpha\beta}(R,t)&=\int\rho_W^{\alpha\beta}(R,P,t)\odif{P} ,\\
					\eta_{\alpha\beta}(P,t)&=\int\rho_W^{\alpha\beta}(R,P,t)\odif{R} .
				\end{align}
			\end{subequations}
			From their definitions, as shown in Appendix~\ref{app:elm marg pos}, it follows that both \(n_\alpha(R,t)\) and \(n_i(R,t)\) should be positive-definite, whereas only \(\eta_i(P,t)\) must be.

			In the next section, we demonstrate both numerically and analytically that the dynamics of a system with an initially positive-definite phase space density can evolve under the QCLE to a pseudo-density with negative regions in phase space. More significantly, the evolved pseudo-density violates the positive-definiteness of the marginal densities in the bath position and momentum. To quantify the violation of positive semi-definiteness of the marginal densities, we define a negativity index \(\mathcal{N}\) for a marginal \(f(X,t)\):
			\begin{equation}\label{eq:neg def}
				\mathcal{N}[f(X,t)]=\begin{dcases}
					0\qc\text{if }f(X,t)=0\text{ everywhere}\\
					\dfrac{\displaystyle \int\big(|f(X,t)|-f(X,t)\big)\odif{X}}{\displaystyle 2\int|f(X,t)|\odif{X}}\qc\text{otherwise}
				\end{dcases}
			\end{equation}
			where \(f(X,t)\) is either \(n(R,t)\) or \(\eta (P,t)\). From this definition, it follows that the negativity index lies in the range \([0,1]\): If the marginal density is positive semi-definite, it has a negativity index of zero, while a higher negativity index indicates that a larger fraction of the range of a relevant variable has a negative marginal pseudo-density.

	\section{\label{sec:negativeMarginals}Negative Marginals Arising in the QCLE}
		Although the QCLE was originally developed to describe the dynamics of a quantum subsystem coupled to a \textit{large} classical bath, it is instructive to analyze its mathematical structure to identify potential violations of physical behavior using simple systems. Such violations can be revealed in numerically exact simulations of low-dimensional models, where neglecting the higher-order Poisson bracket terms in the formally exact evolution of \(\hat{\rho}_W\), given in Eq.~\ref{eq:QLE}, becomes significant. In high-dimensional systems with an extended (often thermal) bath, marginal densities such as \(n(R,t)\) for a classical coordinate \(R\) directly coupled to the quantum subsystem are obtained by integrating over the remaining classical bath coordinates. However, integrating the joint probability density \(n(R,R',t)\), which should always be positive definite, over all the other bath coordinates \(R'\) to obtain the marginal probability \(n(R,t)=\int n(R,R',t)\odif{R'}\) may obscure any physical violations that arise from the QCLE.~For this reason, it is useful to consider simple systems with small classical baths, such as the one-dimensional dual avoided crossing model.

		\subsection{\label{sec:DAC}The Dual Avoided Crossing Model}
			\subsubsection{Model and Final Population}
				Tully's dual avoided crossing (DAC) model\cite{10.1063/1.459170}, originally introduced to investigate nonadiabatic effects in a two-level system coupled to the motion of a massive particle, has become a standard benchmark for testing mixed quantum\textendash{}classical dynamical methods. Owing to its low dimensionality, numerically exact quantum solutions are readily obtainable and can be directly compared to approximate solutions. The model's simplicity permits numerically converged solutions for both the full evolution of \(\rho_W(R,P,t)\), defined in Eq.~\ref{eq:QLE}, and the approximate QCLE (see Eq.~\ref{eq:QCLE}). Simulations of the QCLE initialized with a Gaussian wavepacket show that, while the adiabatic populations and the average position and momentum agree quantitatively with those predicted ny the exact quantum dynamics across a wide range of initial energies, significant discrepancies arise in the phase-space pseudo-densities at low energies. These discrepancies manifest as negativities not only in the pseudo-density\textemdash{}where they might be expected, given their presence in the partial Wigner transform of the exact quantum density matrix\textemdash{}but also in the marginal densities, where positivity should be preserved.

\begin{figure*}
					\includegraphics[keepaspectratio]{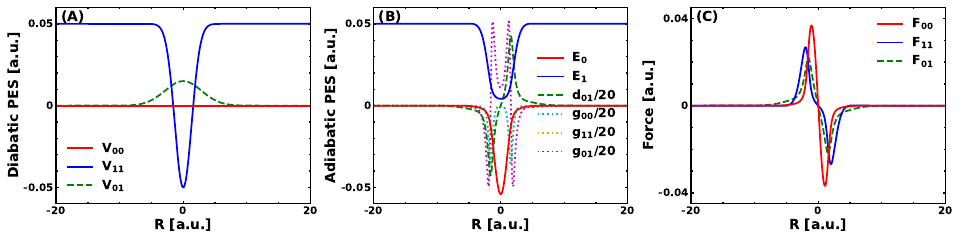}
					\caption{\label{fig:dac pes} Tully's ``Dual Avoided Crossing'' (DAC) model: (a) diabatic potential \(V_{ij}(R)=\braket[3]{i}{V}{j}\) given in Eq.~\ref{eqs:DAC diaV}, (b) adiabatic potential \(E_{\alpha}(R)\) by diagonalization of \(V_{\alpha\beta}(R)\), first order coupling \(d_{\alpha\beta}(R)=\braket*[3]{\alpha(R)}{\pdv{}/{R}}{\beta(R)}\) and second order coupling \(g_{\alpha\beta}(R)=\braket*[3]{\alpha(R)}{\pdv[2]{}/{R}}{\beta(R)}\) (\(g_{00}=g_{11}\) here) in the adiabatic basis, and (c) ``force'' \(F^{\alpha\beta}_W(R)=-\braket*[3]{\alpha(R)}{\pdv{V}/{R}}{\beta(R)}\).}
				\end{figure*}

				The DAC model is a scattering model with two avoided crossing regions, defined by the diabatic potential matrix
				\begin{equation}\label{eqs:DAC diaV}
					\begin{aligned}
						V_{00}(R)&=0\\
						V_{11}(R)&=-A\exp(-Bx^2)+E_0\\
						V_{01}(R)&=V_{10}(R)=C\exp(-Dx^2) ,
					\end{aligned}
				\end{equation}
				with parameters \(E_0=0.05\), \(A=0.1\), \(B=0.28\), \(C=0.015\), and \(D=0.06\). The particle mass is \(M=2000\), and Planck's constant is set to \(\hbar=1\). The matrix elements of the diabatic and adiabatic potentials (\(V\) and \(E\), respectively), nonadiabatic coupling elements (\(1\)st order \(d\) and \(2\)nd order \(g\)), and adiabatic ``force'' (\(F\)) are shown in Fig.~\ref{fig:dac pes}.

				Initially, the wavepacket is fully populated on the adiabatic ground state,
				\begin{equation}
					\Psi(R,0)=\psi_0(R)\ket{0(R)} ,
				\end{equation}
				expressed as a Gaussian centered at a point \(R_0\) with width \(\sigma_R\) and initial momentum \(P_0\)
				\begin{equation}
					\psi_0(R)={(2\pi)}^{-\frac{1}{4}}\sigma_R^{-\frac{1}{2}}\exp\ab(-{\ab(\frac{R-R_0}{2\sigma_R})}^2+\frac{\iu P_0 R}{\hbar})\label{eq:psi_0} ,
				\end{equation}
				and its corresponding adiabatic PWTDM is calculated by the partial Wigner transform scheme in Eqs.~\ref{eqs:WT}.

\begin{figure}[htb]
					\includegraphics[keepaspectratio]{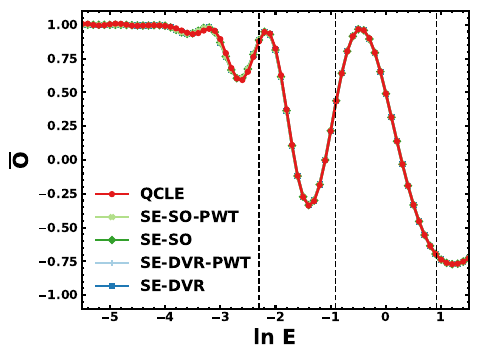}
					\caption{\label{fig:dac ppl} Final asymptotic adiabatic population difference between adiabatic ground and excited states of DAC model as a function of the logarithm of the initial energy. This initial energy is defined from the initial momentum \(P_0\) by \(E=P_0^2/(2M)\). The converged results obtained using the QCLE, the split-operator method for the \sch{} equation (SE-SO) and its PWTDM (SE-SO-PWT), and the DVR method for \sch{} equation by diagonalization of the Hamiltonian (SE-DVR) and its PWTDM (SE-DVR-PWT) are shown as red circles, dark green diamonds, light green crosses, dark blue squares, and light blue plus signs, respectively. The vertical dashed lines correspond to the different initial energy regimes, with \(P_0=20\) a.u., \(40\) a.u., and \(100\) a.u., from left to right in the figure. \(P_0=4\) a.u., with \(\ln E=-5.521\), is at the left edge of the panel.}
				\end{figure}

				To illustrate the asymptotic populations for this two-level system, we introduce the observable\cite{10.1063/1.1433502}
				\begin{equation}
					\hat{O}=\hat{\sigma}_3=\begin{pmatrix}1&0\\0&-1\end{pmatrix} ,
				\end{equation}
				where the matrix elements are expressed in the adiabatic basis. The expectation value \(\langle\hat{O}_W\rangle\) gives the population difference between the two adiabatic surfaces. Fig.~\ref{fig:dac ppl} shows how the population difference between the adiabatic potential energy surfaces, after the system traverses through the region of significant nonadiabatic coupling, depends on the initial energy. All numerical results were calculated using various numerical methods, and their convergence was verified with a discretization error of less than \(0.0026\). These methods include the discrete variable representation (DVR)\cite{10.1063/1.462100} and the split-operator method with fast Fourier transform (FFT)\cite{FEIT1982412, KOSLOFF198335} for the time-dependent \sch{} equation (TDSE), and the split-operator method for QCLE\cite{10.1063/1.478811}. Details of numerical calculations are given in Appendix~\ref{app:numeric method}. For the \sch{} equation results, we also numerically transform the final adiabatic wavefunction to a partial phase-space representation using the partial Wigner transform scheme in Eqs.~\ref{eqs:WT}. As is clear from Fig.~\ref{fig:dac ppl}, the asymptotic final population differences between the adiabatic states computed using the numerically converged solutions of the QCLE and the \sch{} equation match with discrepancies of less than \(0.025\) for all initial values of the energy.

			\subsubsection{\label{sec:DAC PSD}Phase-Space Pseudo-Density}
\begin{figure*}
					\includegraphics[keepaspectratio]{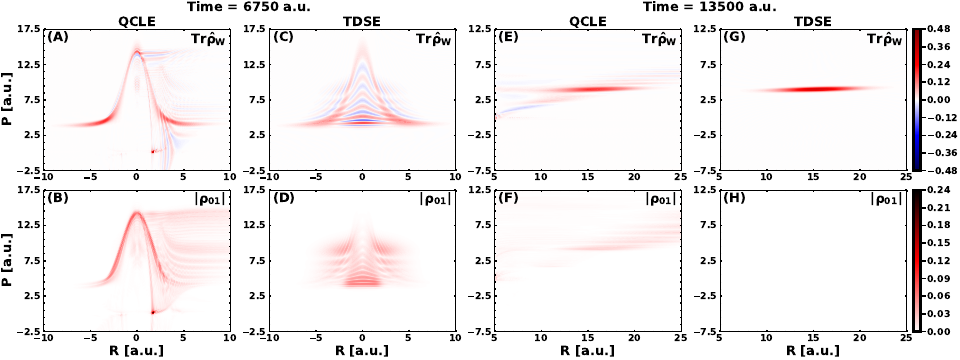}
					\caption{\label{fig:p=4 pwtdm} Color maps of the phase-space pseudo-density \(\rho(R,P,t)=\Tr\hat{\rho}_W\) in the low momentum regime, shown in panels A, C, E, and G, and the norm of the coherence in the adiabatic basis \(|\rho_{W}^{01}|\), shown in panels B, D, F, and H. The initial conditions are \(R_0=-15\) a.u.\ and \(P_0=4\) a.u.\ Panels A, B, E, and F are obtained by evolution of the QCLE, and C, D, G, and H are obtained from the partial Wigner transform of the \sch{} equation. Panels A\textendash{}D and E\textendash{}H correspond to \(t=6750\) a.u.\ and \(t=13500\) a.u., where \(\langle R\rangle\approx0\) and \(\langle R\rangle\approx|R_0|\), respectively.}
				\end{figure*}

\begin{figure*}
					\includegraphics[keepaspectratio]{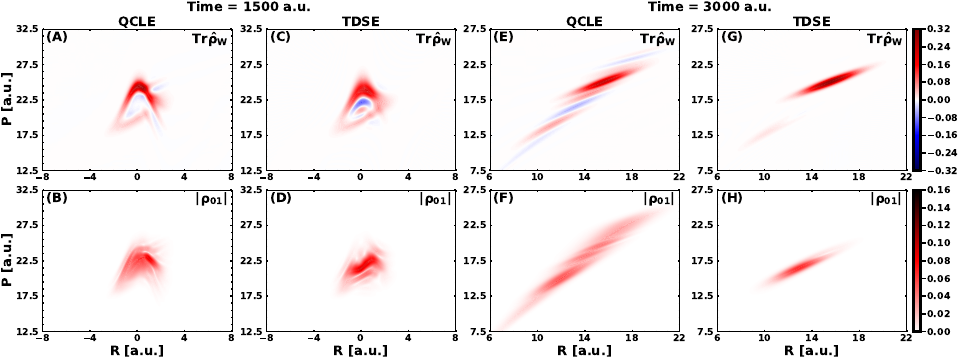}
					\caption{\label{fig:p=20 pwtdm} Color maps of the phase-space pseudo-density \(\rho(R,P,t)\) in the intermediate momentum regime, shown in panels A, C, E, and G, and the norm of the coherence in the adiabatic basis \(|\rho_{W}^{01}|\), shown in panels B, D, F, and H. The initial conditions are \(R_0=-15\) a.u.\ and \(P_0=20\) a.u.\ Panels A, B, E, and F are obtained by evolution of the QCLE, and C, D, G, and H are obtained from the partial Wigner transform of the \sch{} equation. Panels A\textendash{}D and E\textendash{}H correspond to \(t=1500\) a.u.\ and \(t=3000\) a.u., where \(\langle R\rangle\approx0\) and \(\langle R\rangle\approx|R_0|\), respectively.}
				\end{figure*}

\begin{figure*}
					\includegraphics[keepaspectratio]{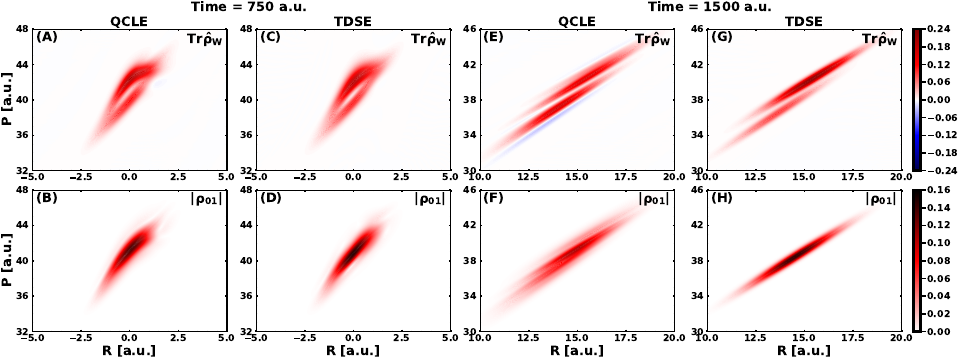}
					\caption{\label{fig:p=40 pwtdm} Color maps of the phase-space pseudo-density \(\rho(R,P,t)\) in the high momentum regime, shown in panels A, C, E, and G, and the norm of the coherence in the adiabatic basis \(|\rho_{W}^{01}|\), shown in panels B, D, F, and H. The initial condition is \(R_0=-15\) a.u.\ and \(P_0=40\) a.u.\ Panels A, B, E, and F are obtained by evolution of the QCLE, and C, D, G, and H are obtained from the partial Wigner transform of the \sch{} equation. Panels A\textendash{}D and E\textendash{}H correspond to \(t=750\) a.u.\ and \(t=1500\) a.u., where \(\langle R\rangle\approx0\) and \(\langle R\rangle\approx|R_0|\), respectively.}
				\end{figure*}

\begin{figure*}
					\includegraphics[keepaspectratio]{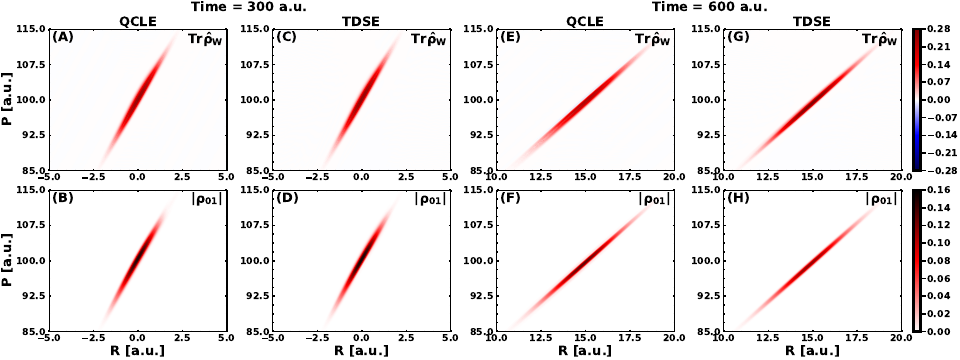}
					\caption{\label{fig:p=100 pwtdm} Color maps of the phase-space pseudo-density \(\rho(R,P,t)\) in the high momentum regime, shown in panels A, C, E, and G, and the norm of the coherence in the adiabatic basis \(|\rho_{W}^{01}|\), shown in panels B, D, F, and H. The initial condition is \(R_0=-15\) a.u.\ and \(P_0=100\) a.u.\ Panels A, B, E, and F are obtained by integrating the QCLE, and C, D, G, and H are obtained from the partial Wigner transform of the \sch{} equation. Panels A\textendash{}D and E\textendash{}H correspond to \(t=300\) a.u.\ and \(t=600\) a.u., where \(\langle R\rangle\approx0\) and \(\langle R\rangle\approx|R_0|\), respectively.}
				\end{figure*}

				Although there is quantitative agreement in the populations of the adiabatic states in the QCLE and quantum dynamics, differences in maps of the pseudo-density of the phase space for the bath degree of freedom \(\rho(R,P,t)\) are apparent, most notably for low initial values of the bath momentum \(P\). To examine these differences in detail, we highlight the phase-space pseudo-densities for four different energy regimes, as indicated in Fig.~\ref{fig:dac ppl}: one with a low initial momentum \(P_0=4\), corresponding to an initial energy below the energy gap \(E_0\) (with \(\ln E=-5.521\) and \(E/E_0=0.08\)) where there is negligible population transfer from the ground to the excited state; a medium-low initial momentum \(P_0=20\) (with \(\ln E=-2.303\) and \(E/E_0=2\)), where destructive interference effects between the adiabatic states reduce population transfer; a medium-high momentum \(P_0=40\) (with \(\ln E=-0.916\) and \(E/E_0=8\)), where the asymptotic final population on the excited state is \(0.292\); and a high initial momentum \(P_0=100\) (with \(\ln E=0.916\) and \(E/E_0=50\)), where the asymtotic final population on the excited state is \(0.852\). In Fig.~\ref{fig:p=4 pwtdm}, Fig.~\ref{fig:p=20 pwtdm}, Fig.~\ref{fig:p=40 pwtdm} and Fig.~\ref{fig:p=100 pwtdm}, color maps of the phase-space pseudo-density \(\rho(R,P,t)\) and the magnitude of the off-diagonal density matrix element at a time when the system evolves through the region with strong nonadiabatic coupling, \(\langle R\rangle\approx0\), and a later time when the system has passed entirely through the coupling region, \(\langle R\rangle\approx|R_0|\), are shown for the respective energy regimes.

				As is evident from comparing the panels in Figs.~\ref{fig:p=4 pwtdm}A and~\ref{fig:p=4 pwtdm}C, major discrepancies in the color maps of the pseudo-density from the QCLE and the full quantum approach occur in the low-momentum regime. In this regime, the partial Wigner transform of the \sch{} equation solutions displays strong tunneling effects that manifest as oscillatory patterns in the color maps over a range of momentum when passing through the spatial regions between the two avoided crossings of the adiabatic potential energy surface. By contrast, the system evolves nearly adiabatically under the QCLE.~The two evolution equations produce different pseudo-density profiles that persist after the wavepacket passes through the coupling region; the negative regions of the phase-space pseudo-density disappear under the \sch{} equation evolution in Fig.~\ref{fig:p=4 pwtdm}G but remain prominent in the QCLE in Fig.~\ref{fig:p=4 pwtdm}E. Although the overall populations agree, the final density of the coherence in Fig.~\ref{fig:p=4 pwtdm}H nearly vanishes everywhere under the \sch{} equation evolution, while the density of the coherence in the QCLE evolution remains (see Fig.~\ref{fig:p=4 pwtdm}F).

				In the intermediate energy regime, \(P_0=20\) a.u., there are significant differences in the phase-space pseudo-densities both when the system moves through the strong coupling region and in the final asymptotic pseudo-densities. This is particularly evident in the coherence densities, which differ in their shape (Figs.~\ref{fig:p=20 pwtdm}B and~\ref{fig:p=20 pwtdm}D), and in the strength of the negativity of the phase-space pseudo-density (Figs.~\ref{fig:p=20 pwtdm}A and~\ref{fig:p=20 pwtdm}C). At long times, the densities of both the classical phase space and the coherence are more dispersed and exhibit more extensive negative regions in the QCLE evolution, even though the final integrated populations of each adiabatic surface agree quantitatively.

				As the initial average energy of the wavepacket increases, the phase space densities of each density matrix element under the QCLE and \sch{} equation evolution progressively resemble one another, as can be seen in Figs.~\ref{fig:p=40 pwtdm} and~\ref{fig:p=100 pwtdm}. One might expect that at high energies, for which the QCLE evolution is virtually indistinguishable from the full quantum evolution, the negative regions of \(\rho(R,P,t)\) would disappear since the bath behaves more classically. However, in Figs~\ref{fig:p=40 pwtdm}E and~\ref{fig:p=100 pwtdm}E, we see that this is not the case. Interestingly, for \(P_0=100\), even though the phase space densities from the QCLE and the \sch{} equation in Fig.~\ref{fig:p=100 pwtdm} are nearly indistinguishable\textemdash{}suggesting that the QCLE is a very good approximation for the quantum evolution\textemdash{}the ``densities'' still retain negative values in regions of phase space. Nonetheless, as the simple low-dimensional model indicates, the QCLE accurately describes the evolution of average properties even when negativities appear, and the negativity of the phase-space pseudo-density for the bath degree of freedom does not indicate a failure of the QCLE.~For a large classical bath, the fraction of phase space where the pseudo-density is negative likely becomes exponentially small.

			\subsubsection{Marginal Density and Negativity}
\begin{figure*}[ht]
					\includegraphics[keepaspectratio]{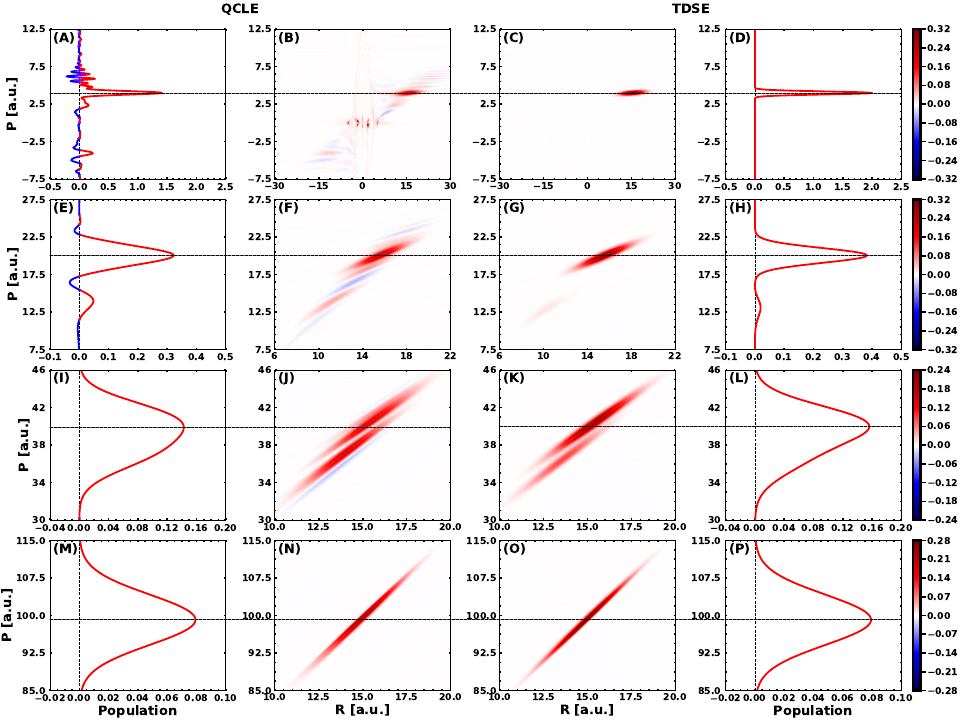}
					\caption{\label{fig:phase and marg} Asymptotic phase-space pseudo-density \(\rho(R,P,t)\) (panels B, C, F, G, J, K, N, and O) and the corresponding momentum marginal densities \(\eta(P,t)\) (panels A, D, E, H, I, L, M, and P), where the negative values are plotted in blue. The momentum corresponding to the maximum marginal density and the zero of the marginal density are indicated by a black dashed line. Panels A, B, E, F, I, J, M, and N are obtained from the evolution of the QCLE, while panels C, D, G, H, K, L, O, and P from the evolution of the \sch{} equation. Panels A\textemdash{}D, E\textemdash{}H, I\textemdash{}L, and M\textemdash{}P correspond to initial momenta of \(P_0=4\) a.u., \(P_0=20\) a.u., \(P_0=40\) a.u., and \(P_0=100\) a.u., respectively.}
				\end{figure*}

				Unlike the quasi-probabilistic phase-space pseudo-density, the marginal densities of bath variables, defined in Eqs.~\ref{eqs:marginals}, should remain positive at all times and for all bath positions and momenta. However, the marginals under the QCLE evolution, obtained by numerical integration of the phase-space pseudo-density \(\rho(R,P,t)\), clearly violate positivity in the low- and intermediate-energy regimes, as shown in Fig.~\ref{fig:phase and marg}. The marginal distribution, though physically meaningful, is less informative than the full phase-space distribution, and the marginals obtained from the QCLE quickly converge to those obtained from the \sch{} equation and become positive definite. Note that at high initial energies, such as \(P_0=100\) a.u., a small negative region appears in the QCLE pseudo-density (Fig.~\ref{fig:phase and marg}N), unlike the \sch{} equation pseudo-densities (Fig.~\ref{fig:phase and marg}O), whereas the marginal distributions (Fig.~\ref{fig:phase and marg}M and P) remain indistinguishable.

\begin{figure}[htbp]
					\includegraphics[keepaspectratio]{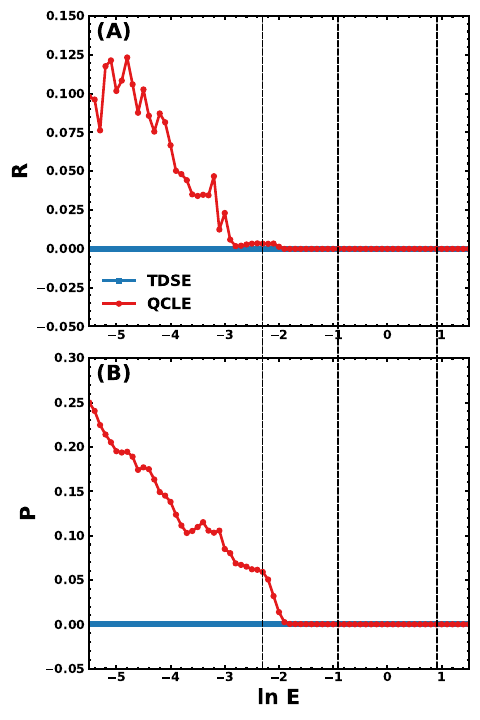}
					\caption{\label{fig:dac neg} Asymptotic negativity index in the DAC model: (a) overall marginal of \(R\), \(\mathcal{N}[n(R,t)]\), calculated with a spatial grid spacing of \(\Delta R'=0.015\) and a momentum grid spacing of \(k=2\) relative to the thermal de Broglie wavelength, and (b) overall marginal of \(P\), \(\mathcal{N}[\eta(P,t)]\), calculated with a grid spacing of \(\Delta R'=0.05\) and \(k=5\), with respect to the logarithm of the initial energy. This energy is related to the initial momentum \(P_0\) by \(E=P_0^2/(2M)\). The result of the QCLE and the PWTDM of the time-dependent \sch{} equation are shown as red circles and blue squares, respectively. Details of the grid spacing and grid discretization error are discussed in Apprendix~\ref{app:numeric method}. The vertical dashed lines correspond to the different initial energy regimes, with \(P_0=20\) a.u., \(40\) a.u., and \(100\) a.u., from left to right in the figure. \(P_0=4\) a.u.\ is out of the left border.}
				\end{figure}

				In Fig.~\ref{fig:dac neg}, the asymptotic values of the negativity index, defined in Eq.~\ref{eq:neg def}, for the spatial and momentum marginals are shown as functions of the initial energy. As expected, the numerical solutions for the negativity index of the position and momentum marginals of the \sch{} equation vanish for all initial conditions, as they must. This numerically confirms that the overall marginal densities of position and momentum (\(n(R,t)\) and \(\eta(P,t)\), respectively), derived from the partial Wigner transform of the quantum density matrix evolving under the exact von Neumann equation, are non-negative. However, the QCLE breaks the positivity semidefiniteness of the marginal density of both the position and the momentum at low initial energies, a non-physical result that indicates a failure of the QCLE to properly describe the evolution of the overall system even though the relative errors of some observables, such as the adiabatic populations or the average position and momentum of classical degrees of freedom, remain small. Compared with expectation values of bath or quantum degrees of freedom, the negativity index is more sensitive to discretization error, and different spacings for the \(R\) and \(P\) coordinates are adopted for the sake of computational efficiency. The results for the negativity index at low initial energies in Fig.~\ref{fig:dac neg} are converged and reproducible, with a relative discretization error below \(1\%\). The rapid changes in the negativity index as the initial energy decreases, particularly for the spatial marginal, indicate that intricate quantum effects alter the structure of the bath phase-space pseudo-density at low energies. Note that the initial energy gap is \(E_0=0.05\) a.u.\ for this model, while the minimal initial energy where the negativity index vanishes is around \(\e{-1.9}\approx0.15\) a.u., corresponding to an initial average momentum of \(P_0\approx 24.5\) a.u., roughly three times \(E_0\). Hence, the range of ``low'' energy where the QCLE is expected to fail is wider than might be anticipated on the grounds of other physical observables. On the other hand, the disappearance of negativity in the marginal density\textemdash{}rather than the existence of negativities in the full phase-space density\textemdash{}can serve as a weak criterion for evaluating the validity of QCLE.\@

				Generally speaking, the evolution of the marginal density under the QCLE is complicated to analyze, although general nonlocal equations of motion for the marginal densities can be derived using projection operator methods, as discussed in Appendix~\ref{app:marg EOM}. In the next section, we introduce a simple model that illustrates the emergence of negativity in the marginal density under the QCLE evolution.

		\subsection{\label{sec:const model}Analytical study of the negativity index: The case of constant nonadiabatic coupling}
			A simple model that mimics the passage of a quantum wavepacket through a region of strong nonadiabatic coupling can be constructed to shed light on how negativities in the marginal densities arise in the QCLE.~The simplicity of the model enables local equations of motion for the marginal densitis to be written and analytically solved for both the full quantum and the QCLE dynamics.

\begin{figure*}[htb]
				\includegraphics[keepaspectratio]{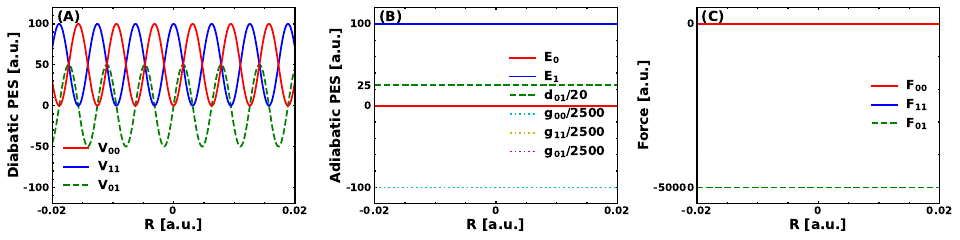}
				\caption{\label{fig:const pes} The constant model: (a) diabatic potential \(V_{\alpha\beta}(R)=\braket[3]{\alpha}{V}{\beta}\) given in Eqs.~\ref{eq:const diaV}, (b) adiabatic potential \(E_{\alpha}(R)=E\delta_{1\alpha}\), first order coupling \(d_{\alpha\beta}(R)=D(\delta_{0\alpha}\delta_{1\beta}-\delta_{1\alpha}\delta_{0\beta})\), second order coupling \(g_{\alpha\beta}(R)=-D^2\delta_{\alpha\beta}\), and (c) ``force'' \(F^{\alpha\beta}_W(R)=-DE(1-\delta_{\alpha\beta})\) in the adiabatic basis. The plot is drawn with \(D=500\) and \(E=100\), for which the approximations in Eq.~\ref{eq:approx qcle cond} hold.}
			\end{figure*}

			We consider a one-dimensional, two-level system for a particle of mass \(M\) evolving in the presence of both a constant adiabatic potential,
			\begin{equation}
				V^{\ad}(R)=\begin{pmatrix}0&0\\0&E\end{pmatrix} ,
			\end{equation}
			and a constant nonadiabatic coupling matrix,
			\begin{equation}
				D^{\ad}(R)=\begin{pmatrix}0&D\\-D&0\end{pmatrix} ,
			\end{equation}
			where \(D\) and \(E\) are \(R\)-independent constants. The corresponding underlying diabatic potential for this model is
			\begin{equation}\label{eq:const diaV}
				V^\dia(R)=\frac{E}{2}I+\frac{E}{2}\begin{pmatrix}
					-\cos(2DR)&\sin(2DR)\\\sin(2DR)&\cos(2DR)
				\end{pmatrix} ,
			\end{equation}
			and the basis transform matrix converting from the adiabatic to the diabatic basis is
			\begin{equation}
				U(R)=\begin{pmatrix}\cos(DR)&\sin(DR)\\-\sin(DR)&\cos(DR)\end{pmatrix} .
			\end{equation}
			In the diabatic representation, the model exhibits a sinusoidal potential. In the adiabatic basis, the adiabatic dynamics of the particle are force-free, while the coupling between adiabatic states induces population and momentum transfer. Features of the model are illustrated in Fig.~\ref{fig:const pes}.

			For this model, the QCLE represented in the adiabatic basis simplifies to
			\begin{subequations}\label{eqs:const model qcle}
				\begin{align}
					\pdv{\rho_W^{00}}{t}=&\frac{DE}{2}\pdv*{\bigg(\rho_W^{01}+\rho_W^{10}\bigg)}{P}-\frac{DP}{M}\bigg(\rho_W^{01}+\rho_W^{10}\bigg)-\frac{P}{M}\pdv{\rho_W^{00}}{R} ,\\
					\pdv{\rho_W^{10}}{t}=&\frac{DE}{2}\pdv*{\bigg(\rho_W^{00}+\rho_W^{11}\bigg)}{P}+\frac{DP}{M}\bigg(\rho_W^{00}-\rho_W^{11}\bigg)-\ab(\frac{\iu E}{\hbar}+\frac{P}{M}\pdv{}{R})\rho_W^{10} ,\\
					\pdv{\rho_W^{11}}{t}=&\frac{DE}{2}\pdv*{\bigg(\rho_W^{01}+\rho_W^{10}\bigg)}{P}+\frac{DP}{M}\bigg(\rho_W^{01}+\rho_W^{10}\bigg)-\frac{P}{M}\pdv{\rho_W^{11}}{R} ,
				\end{align}
			\end{subequations}
			and, accordingly, the evolution equation for the momentum marginal densities is
			\begin{subequations}\label{eqs:const model qcle p marg evo}
				\begin{align}
					\pdv{\eta_0(P,t)}{t}&=-\frac{2DP}{M}\eta_r+DE\pdv{\eta_r}{P} ,\\
					\pdv{\eta_r(P,t)}{t}&=\frac{DP}{M}(\eta_0-\eta_1)+\frac{DE}{2}\pdv*{(\eta_0+\eta_1)}{P}+\frac{E}{\hbar}\eta_i ,\\
					\pdv{\eta_i(P,t)}{t}&=-\frac{E}{\hbar}\eta_r ,\\
					\pdv{\eta_1(P,t)}{t}&=\frac{2DP}{M}\eta_r+DE\pdv{\eta_r}{P} ,
				\end{align}
			\end{subequations}
			where \(\eta_r=\Re \eta_{10}\) and \(\eta_i=\Im \eta_{10}\).

			We consider an initial adiabatic wavefunction of the form,
			\begin{equation}
				\Psi(R,0)=\psi_0(R)\Big(\cos\theta\ket{0(R)}+\sin\theta\ket{1(R)}\Big), \label{eq:const init wfn}
			\end{equation}
			where \(\psi_0(R)\) is a Gaussian wavepacket of width \(\sigma_R\) centered at \(R_0\) with momentum \(P_0\), as in Eq.~\ref{eq:psi_0}, and \(\theta\in[0,2\pi)\) defines the mixing of the ground and excited adiabatic states. Defining the normal density \(N(X;X_0;\sigma_X)\), 
			\begin{equation}
				N(X;X_0;\sigma_X)=\frac{1}{\sqrt{2\pi}\sigma_X}\exp\ab(-\frac{1}{2}{\ab(\frac{X-X_0}{\sigma_X})}^2) ,
			\end{equation}
			the corresponding partial Wigner-transformed adiabatic density matrix elements are
			\begin{subequations}
				\begin{align}
					\rho_W^{00}(R,P,0)&=\frac{N(R;R_0;\sigma_R)}{4}\bigg(2\cos(2\theta)N(P;P_0;\sigma_P)\nonumber\\
					&\qquad+N(P;P_0+\hbar D;\sigma_P)+N(P;P_0-\hbar D;\sigma_P)\bigg) ,\\
					\rho_W^{10}(R,P,0)&=\frac{N(R;R_0;\sigma_R)}{4}\bigg(2\sin(2\theta)N(P;P_0;\sigma_P)\nonumber\\
					&\qquad-\iu\big(N(P;P_0+\hbar D;\sigma_P)+N(P;P_0-\hbar D;\sigma_P)\big)\bigg) ,\\
					\rho_W^{11}(R,P,0)&=\frac{N(R;R_0;\sigma_R)}{4}\bigg(-2\cos(2\theta)N(P;P_0;\sigma_P)\nonumber\\
					&\qquad+N(P;P_0+\hbar D;\sigma_P)+N(P;P_0-\hbar D;\sigma_P)\bigg) .
				\end{align}
			\end{subequations}
			and the corresponding initial marginal densities of momentum are
			\begin{subequations}\label{eqs:const model init P marg}
				\begin{align}
					\eta_0(P,0)&=\frac{1}{4}\bigg(2\cos(2\theta)N(P;P_0;\sigma_P)+N(P;P_0+\hbar D;\sigma_P)+N(P;P_0-\hbar D;\sigma_P)\bigg) ,\\
					\eta_r(P,0)&=\frac{\sin(2\theta)}{2} N(P;P_0;\sigma_P) ,\\
					\eta_i(P,0)&=-\frac{1}{4}\bigg(N(P;P_0+\hbar D;\sigma_P)+N(P;P_0-\hbar D;\sigma_P)\bigg) ,\\
					\eta_1(P,0)&=\frac{1}{4}\bigg(-2\cos(2\theta)N(P;P_0;\sigma_P)+N(P;P_0+\hbar D;\sigma_P)+N(P;P_0-\hbar D;\sigma_P)\bigg) .
				\end{align}
			\end{subequations}
			Note that although the initial total marginal density \(\eta(P,0)=\eta_0(P,0)+\eta_1(P,0)\) is positive everywhere, the initial marginal density \(\eta_1(P,0)\) can be negative if \(\sigma_P\) is small and only the ground adiabatic state is initially populated (\(\theta=0\)). This negativity is not forbidden, since the distribution of momentum in a specific adiabatic state is not observable, as discussed in Appendix~\ref{app:elm marg pos}.

			\begin{figure}[tb]
				\includegraphics[keepaspectratio]{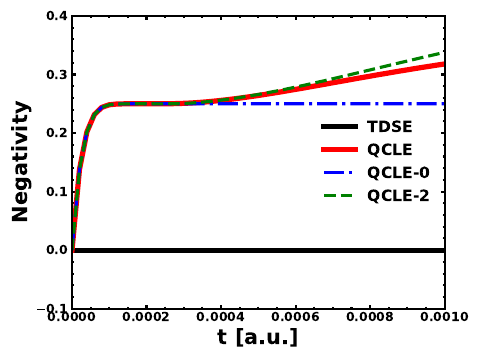}
				\caption{\label{fig:const neg} Numerical and predicted negativity index of the marginal momentum density \(\mathcal{N}[\eta(P,t)]\) versus time for the constant model with \(D=500\), \(E=100\), \(M=200\), \(P_0=20\), and \(\sigma_P=2\). Black solid, red solid, green dashed, and blue dash-dotted lines correspond to the analytical solutions of the \sch{} equation, the numerical solution of the QCLE, and the different orders of the perturbative solutions of the QCLE, respectively.}
			\end{figure}

			Given that the initial marginals are linear combinations of normal distributions (Gaussians), we would expect from the form of the equations of the marginals, Eqs.~\ref{eqs:marginals}, that the marginals remain linear combinations of Gaussians for short times. Noting that the maximum values of terms in the evolution equations can be estimated as,
			\begin{subequations}
				\begin{align}
					\max_P \frac{DP}{M}N(P;P_0;\sigma_P)&\approx\frac{D|P_0|}{\sqrt{2\pi}\e{}\sigma_P M} ,\\
					\max_P \frac{E}{\hbar}N(P;P_0;\sigma_P)&=\frac{E}{\sqrt{2\pi}\sigma_P\hbar} ,\\
					\max_P DE\pdv{}{P}N(P;P_0;\sigma_P)&=\frac{DE}{\sqrt{2\pi}\e{}\sigma_P^2} ,
				\end{align}
			\end{subequations}
			and if
			\begin{equation}\label{eq:approx qcle cond}
				\frac{\hbar D}{\sigma_P} \gg \e{},\frac{\hbar D|P_0|}{EM} ,
			\end{equation}
			the derivative term with respect to the momentum dominates in Eqs.~\ref{eqs:const model qcle p marg evo}, and the evolution of the marginal densities can be approximated as
			\begin{subequations}
				\begin{align}
					\pdv{\eta_0(P,t)}{t}&=DE\pdv{\eta_r}{P} ,\\
					\pdv{\eta_r(P,t)}{t}&=\frac{DE}{2}\pdv*{(\eta_0+\eta_1)}{P} ,\\
					\pdv{\eta_i(P,t)}{t}&=0 ,\\
					\pdv{\eta_1(P,t)}{t}&=DE\pdv{\eta_r}{P} ,
				\end{align}
			\end{subequations}
			whose solution gives
			\begin{multline}\label{eq:const model large D small sigma sol}
				\eta(P,t)=\frac{1}{2}\bigg(\eta_0(P+DEt,0)+\eta_1(P+DEt,0)+\eta_0(P-DEt,0)\\
				+\eta_1(P-DEt,0)+2\eta_r(P+DEt,0)-2\eta_r(P-DEt,0)\bigg) .
			\end{multline}
			Thus, by taking an equal initial mixture of the ground and excited adiabatic states (\(\theta=\pi/4\)), we have
			\begin{multline}\label{eq:const P marg QCLE pert sol}
				\eta(P,t)=\frac{1}{4}\bigg(2N(P;P_0-DEt;\sigma_P)-2N(P;P_0+DEt;\sigma_P)+N(P;P_0+\hbar D+DEt;\sigma_P)\\
				+N(P;P_0-\hbar D+DEt;\sigma_P)+N(P;P_0+\hbar D-DEt;\sigma_P)+N(P;P_0-\hbar D-DEt;\sigma_P)\bigg) ,
			\end{multline}
			where \(\eta(P,t)<0\) is negative in the vicinity of \(P_0+DEt\) when the Gaussian contributions to \(\eta\) are well-separated. This scenario will arise when the nonadiabatic coupling strength \(D\), mass \(M\), and energy gap \(E\) are large and the initial momentum \(P_0\) is small and narrowly distributed to satisfy the conditions
			\begin{equation}
				\frac{\hbar D}{\sigma_P}\gg e,\qand \frac{\sigma_P P_0}{EM}\ll1 .
			\end{equation}

			\begin{figure}
				\includegraphics[keepaspectratio]{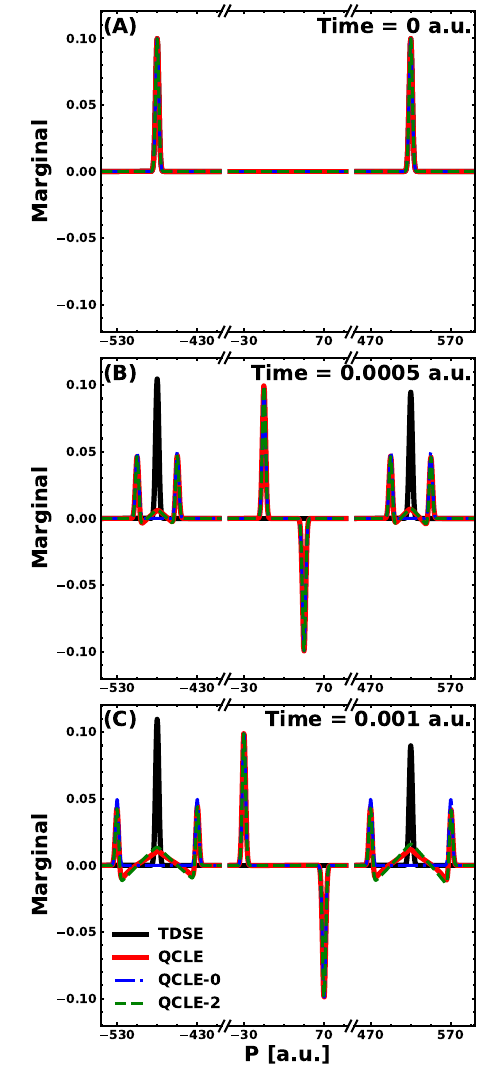}
				\caption{\label{fig:const marg p} Numerical and predicted marginal density of momentum \(\eta(P,t)\) for the constant model with \(D=500\), \(E=100\), \(M=200\), \(P_0=20\), and \(\sigma_P=2\) at three different times. The line colors are the same as in Fig.~\ref{fig:const neg}. For the cut-out range of momentum, the marginal density is vanishingly small. Panels A\textemdash{}C correspond to \(t=0\), \(5.0\times 10^{-4}\) a.u., and \(1.0\times 10^{-3}\) a.u., respectively.}
			\end{figure}

			Numerical results support the analysis above. Choosing parameters \(D=500\), \(E=100\), \(M=200\), \(P_0=20\), and \(\sigma_P=2\) so that \(\hbar D/\sigma_P=250\) and \(\sigma_P P_0/(EM)=2.0\times 10^{-3}\), grid solutions and the predicted form of the marginal densities agree for the QCLE and \sch{} equation dynamics (see Appendix~\ref{app:const TDSE sol} for a discussion of the solution of these equations). As is evident in Figs.~\ref{fig:const neg} and~\ref{fig:const marg p}, Eq.~\ref{eq:const P marg QCLE pert sol} is a good approximate solution of the QCLE.~Higher-order perturbation corrections (see Appendix~\ref{app:const pert sol}) improve the solution at the cost of higher complexity. The correction terms reproduce the peak and the adjacent valleys at \(P=-480\) a.u.\ and \(P=520\) a.u.\ that Eq.~\ref{eq:const P marg QCLE pert sol} does not predict, and the predicted negativity index better matches the numerical solution of the model.

\begin{figure}
				\includegraphics[keepaspectratio]{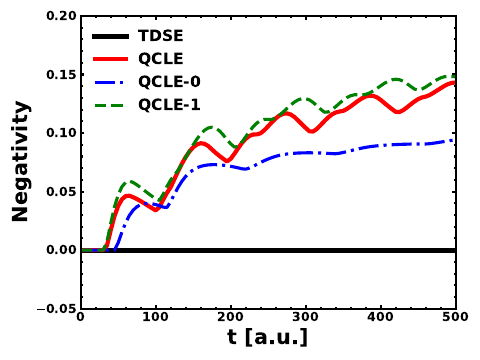}
				\caption{\label{fig:const as dac neg} Numerical and predicted negativity index of the marginal momentum density \(\mathcal{N}[\eta(P,t)]\) versus time for the constant model with \(D=1\), \(E=0.05\), \(M=2000\), \(P_0=20\), and \(\sigma_P=1\). The line colors are the same as in Fig.~\ref{fig:const neg}.}
			\end{figure}

\begin{figure}
				\includegraphics[keepaspectratio]{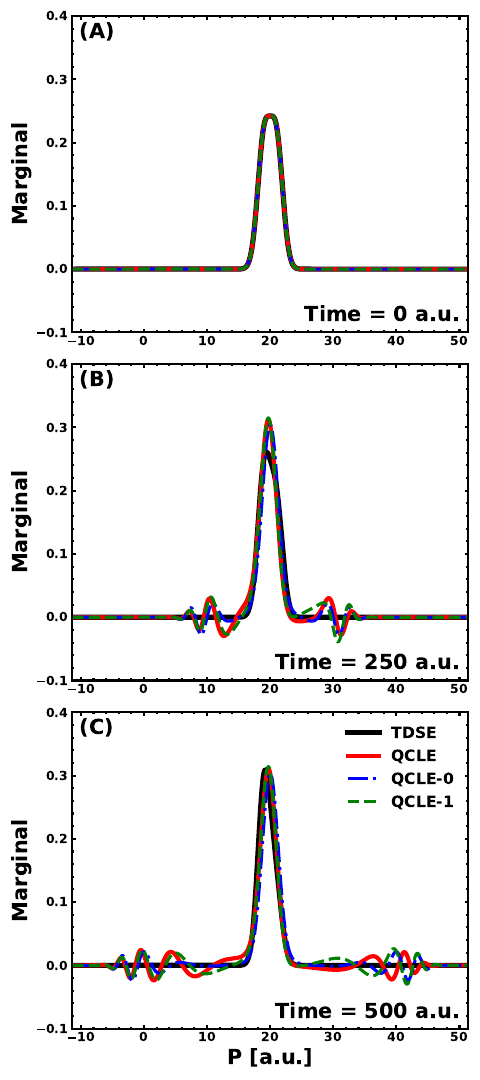}
				\caption{\label{fig:const as dac marg p} Numerical and predicted marginal momentum density \(\eta(P,t)\) for the constant model with \(D=1\), \(E=0.05\), \(M=2000\), \(P_0=20\), and \(\sigma_P=1\) at three different times. The line colors are the same as in Fig.~\ref{fig:const neg}. Panels A\textemdash{}C correspond to \(t=0\), \(250\) a.u., and \(500\) a.u., respectively.}
			\end{figure}

			For initial population entirely in the adiabatic ground state, with parameters \(D=1\), \(E=0.05\), \(M=2000\), \(P_0=20\), and \(\sigma_P=1\), similar to the conditions studied in the DAC model, \(\mathcal{N}[\eta(P,t)]>0\) still arises, though \(\eta(P,t)\) deviates from the form predicted in Eq.~\ref{eq:const P marg QCLE pert sol} and the oscillations in the marginal momentum density resemble those observed in the DAC model in the intermediate-energy regime (see Fig.~\ref{fig:phase and marg}). The negativity index and the marginal momentum density at selected times are presented in Figs.~\ref{fig:const as dac neg} and~\ref{fig:const as dac marg p}, respectively. The perturbative solution for the marginal momentum density for this choice of parameters is more involved and is also presented in Appendix~\ref{app:const pert sol}.

		\subsection{\label{sec:analysis-coupling}Evolution equations for the constant model}
			We have shown that the equation of motion for the adiabatic matrix elements following from the QCLE for the constant model is given by Eqs.~\ref{eqs:const model qcle}. In Appendix~\ref{app:const model QCLE vs QLE}, it is shown that the full dynamical equation, Eq.~\ref{eq:QLE}, with higher-order powers of the Poisson bracket operator \(\Lambda\), can also be manipulated for this simple model to yield the exact equations,
			\begin{subequations}\label{eqs:const model exact}
				\begin{align}
					\pdv{\rho_W^{00}(R,P,t)}{t}&=\frac{E}{4\hbar}\bigg(\rho_W^{01}(R,P+\hbar D,t)+\rho_W^{10}(R,P+\hbar D,t)-\rho_W^{01}(R,P-\hbar D,t)-\rho_W^{10}(R,P-\hbar D,t)\bigg)\nonumber\\
					&\quad-\frac{DP}{M}\bigg(\rho_W^{01}(R,P,t)+\rho_W^{10}(R,P,t)\bigg)-\frac{P}{M}\pdv{\rho_W^{00}(R,P,t)}{R}\\
					\pdv{\rho_W^{11}(R,P,t)}{t}&=\frac{E}{4\hbar}\bigg(\rho_W^{01}(R,P+\hbar D,t)+\rho_W^{10}(R,P+\hbar D,t)-\rho_W^{01}(R,P-\hbar D,t)-\rho_W^{10}(R,P-\hbar D,t)\bigg)\nonumber\\
					&\quad+\frac{DP}{M}\bigg(\rho_W^{01}(R,P,t)+\rho_W^{10}(R,P,t)\bigg)-\frac{P}{M}\pdv{\rho_W^{11}(R,P,t)}{R}\\
					\pdv{\rho_W^{10}(R,P,t)}{t}&=\frac{E}{4\hbar}\bigg(\rho_W^{00}(R,P+\hbar D,t)+\rho_W^{11}(R,P+\hbar D,t)-\rho_W^{00}(R,P-\hbar D,t)-\rho_W^{11}(R,P-\hbar D,t)\bigg)\nonumber \\
					&\quad+\frac{DP}{M}\bigg(\rho_W^{00}(R,P,t)-\rho_W^{11}(R,P,t)\bigg)-\frac{P}{M}\pdv{\rho_W^{10}(R,P,t)}{R}\nonumber\\
					&\quad-\frac{\iu}{2}\bigg(\rho_W^{10}(R,P+\hbar D,t)+\rho_W^{10}(R,P-\hbar D,t)\bigg) .
				\end{align}
			\end{subequations}

			By taking \(\sigma_P\) as the characteristic magnitude of the momentum dispersion, comparison of the two sets of equations shows that, while the exact expression is nonlocal in momentum \(P\), the QCLE provides a local approximation by expanding the matrix elements \(\rho_W^{\alpha\beta}(R, P\pm\hbar D,t)\) in Eq.~\ref{eqs:const model exact} to the first order in \(\hbar D/\sigma_P\). The resulting negativities in the constant model arise from replacing the symmetric finite-difference derivative with a local momentum derivative, and the symmetric average with a local term,
			\begin{subequations}
				\begin{align}
					\frac{1}{2}\bigg(\rho_W^{\alpha\beta}(R,P+\hbar D,t)-\rho_W^{\alpha\beta}(R,P-\hbar D,t)\bigg)&=\hbar D\pdv{\rho_W^{\alpha\beta}(R,P,t)}{P}+\mathcal{O}\ab({\ab(\frac{\hbar D}{\sigma_P})}^2)\\
					\frac{1}{2}\bigg(\rho_W^{10}(R,P+\hbar D,t)+\rho_W^{10}(R,P-\hbar D,t)\bigg)&=\rho_W^{10}(R,P,t)+\mathcal{O}\ab({\ab(\frac{\hbar D}{\sigma_P})}^2) ,
				\end{align}
			\end{subequations}
			which amounts to neglecting higher-order Poisson bracket terms \(\Lambda\) in the exact equation. Because each power of \(\Lambda\) scales with \(\hbar D/\sigma_P\), a small parameter outside the coupling region, higher-order terms become significant mainly when the nonadiabatic coupling element is comparable to the inverse length scale of the spatial dispersion of the wavepacket. These results highlight the role of nonlocal momentum terms in maintaining the positivity of phase-space densities and their marginals.

	\section{\label{sec:conc}Discussion and Conclusion}
		In this work, we examine the similarities and differences in both the phase-space pseudo-density and the density of off-diagonal coherences for a pure quantum state evolved under the QCLE and the time-dependent \sch{} equation. The differences in the pseudo-densities under these evolution equations are studied as functions of the initial energy. Our analysis reveals that, for certain low-dimensional models, the QCLE does not preserve the positivity of the marginal densities of bath coordinates\textemdash{}a property that holds rigorously under full quantum evolution. Importantly, the negativities observed in the QCLE are distinct from those arising in other approximate treatments of open quantum system dynamics. Although the QCLE provides an approximation to full quantum evolution, the mixed quantum\textendash{}classical Liouville operator \(\iu\hat{\mathcal{L}}\) is anti-Hermitian, ensuring exact conservation of probability, phase-space volume, energy, and purity. Thus, the loss of positivity is not a result of the introduction of irreversibility through a weak-coupling or a Markovian approximation. To quantify the extent of this violation, we introduce a negativity index, which measures deviations from the positive-definite condition. We find that, as the initial energy increases relative to the subsystem energy gap, differences between the phase-space densities diminish and the negativity index approaches zero.

		It is well known that the partial Wigner transform introduces negativities into the matrix elements \(\rho_W^{\alpha \beta}\) of the density matrix, as does the full Wigner transform in general phase-space formulations of quantum mechanics\cite{PhysRev.40.749}. We find that negative regions in the trace of the transformed density matrix\textemdash{}the classical phase-space pseudo-density \(\rho(R,P,t)\)\textemdash{}persist under the QCLE evolution after the system traverses a nonadiabatic coupling region, even at relatively high initial energies. The appearance of negative values in \(\rho(R,P,t)\) might suggest a breakdown of a mixed quantum\textendash{}classical description, since classical mechanics allows simultaneous specification of position and momentum. However, despite these negativities, the QCLE and exact quantum phase-space densities agree quantitatively in this high-energy regime. In such cases, the QCLE accurately reproduces the expectation value of any operator \(\hat{A}_W(R,P)\) at all times. Notably, the negativity index vanishes at a lower threshold energy, coinciding with the convergence of the QCLE density to the exact quantum result, and may thus serve as a diagnostic for the validity of mixed quantum\textendash{}classical dynamics. Evaluating the negativity index for a particular classical degree of freedom requires reconstructing its marginal density, a challenging but feasible task using sampling algorithms combined with kernel density estimation\cite{10.1214/aoms/1177728190, 10.1214/aoms/1177704472}. Work in this direction is ongoing.

		The physical origin of a nonzero negativity index in QCLE dynamics remains unclear, as it is highly model-dependent. For instance, the QCLE is exact for the spin-boson model\cite{10.1063/1.1433502}, which consists of a harmonic bath bilinearly coupled to a two-level subsystem; consequently, the negativity index vanishes identically in this case. To illustrate how negativities may arise in more general systems in which a wavepacket passes through regions of strong nonadiabaticity, we consider a simple one-dimensional, force-free model with a constant nonadiabatic coupling element. In this model, the QCLE and exact quantum evolution equations for the adiabatic-basis density matrix elements can be compared directly. This comparison shows that the QCLE corresponds to a local equation in classical phase space, whereas the exact evolution of \(\rho_W(R,P,t)\) is nonlocal in momentum, with the degree of nonlocality determined by the ratio of the nonadiabatic coupling element to the dispersion of the momentum.

		Although intriguing, the failure of the QCLE to preserve the positivity of classical marginals is likely to be of limited practical concern for systems where mixed quantum\textendash{}classical dynamics are intended to apply. Our analysis is restricted to low-dimensional quantum scattering models that permit exact numerical propagation of both the full quantum and QCLE equations. Such models differ substantially from condensed-phase systems, where quantum processes occur in a high-dimensional classical environment. In realistic quantum\textendash{}classical systems, the subsystem typically couples to only a few classical coordinates, or the nonadiabatic coupling vector projects significantly onto a small subset of them. Although joint and marginal densities of these coupled coordinates may exhibit pathologies, the fractional phase-space volume supporting negative pseudo-density values is expected to decrease exponentially with the number of classical degrees of freedom, rendering them of little physical importance. It remains an open question whether forbidden negativities\textemdash{}either in the joint probabilities of position or momentum or in their marginals\textemdash{}would still emerge from the QCLE when the number of classical degrees of freedom for the surroundings becomes progressively larger.

	\begin{acknowledgments}
		This work was supported in part by a grant from the Natural Sciences and Engineering Research Council of Canada. Computations were performed on the GPC supercomputer at the SciNet HPC Consortium. SciNet is funded by the Canada Foundation for Innovation under the auspices of Compute Canada, the Government of Ontario, the Ontario Research Fund: Research Excellence, and the University of Toronto\cite{10.1145/3332186.3332195, Loken2010}.
	\end{acknowledgments}

	\section*{Author Declaration}
		\subsection*{Conflict of Interest}
			The authors have no conflicts to disclose.

		\subsection*{Author Contributions}
			Kai Gu: Conceptualization (equal); Data curation; Formal analysis (equal); Investigation (equal); Methodology (equal); Software; Validation (equal); Visualization; Writing \textendash{} original draft (equal); Writing \textendash{} review \& editing (equal). Jeremy Schofield: Conceptualization (equal); Formal analysis (equal); Funding acquisition; Investigation (equal); Methodology (equal); Project administration; Resources; Supervision; Validation (equal); Writing \textendash{} original draft (equal); Writing \textendash{} review \& editing (equal).

	\section*{Data Availability}
		The data that supports the findings of this study are available from the corresponding author upon reasonable request.

\appendix
	\section{\label{app:QCLE conserve}Conserved Quantities under the QCLE evolution}
		In this Appendix, we establish that the total population of quantum states, the expectation value of the energy, and the purity are constants of motion under the QCLE.\@

		\subsection{Population}
			The time derivative of the population is given by
			\begin{align}
				\pdv{\ev{1}}{t}&=\pdv{}{t}\Tr\int\hat{\rho}_W(R,P,t)\odif{R,P}\nonumber\\
				&=\Tr\int\pdv{\hat{\rho}_W}{t}\odif{R,P}=-\Tr\int\iu\hat{\mathcal{L}}\hat{\rho}_W(t)\odif{R,P}\nonumber \\
				&=-\frac{\iu}{\hbar}\int\Tr[\hat{H}_W,\hat{\rho}_W]\odif{R,P}+\frac{1}{2}\Tr\int(\{\hat{H}_W,\hat{\rho}_W\}-\{\hat{\rho}_W,\hat{H}_W\})\odif{R,P}
			\end{align}
			The trace of the commutator term vanishes due to the cyclic property of the trace, and the Poisson bracket terms are
			\begin{align}
				\pdv{\ev{1}}{t}&=\frac{1}{2}\Tr\int\ab(\pdv{\hat{H}_W}{R}\pdv{\hat{\rho}_W}{P}+\pdv{\hat{\rho}_W}{P}\pdv{\hat{H}_W}{R}-\pdv{\hat{H}_W}{P}\pdv{\hat{\rho}_W}{R}-\pdv{\hat{\rho}_W}{R}\pdv{\hat{H}_W}{P})\odif{R,P}\nonumber\\
				&=\frac{1}{2}\Tr\int\odif{R}\eval{\ab(\pdv{\hat{H}_W}{R}\hat{\rho}_W+\hat{\rho}_W\pdv{\hat{H}_W}{R})}_{P=-\infty}^{P=+\infty}-\frac{1}{2}\Tr\int\ab(\pdv{\hat{H}_W}{R,P}\hat{\rho}_W+\hat{\rho}_W\pdv{\hat{H}_W}{R,P})\odif{R,P}\nonumber\\
				&\qquad-\frac{1}{2}\Tr\int\odif{P}\eval{\ab(\pdv{\hat{H}_W}{P}\hat{\rho}_W+\hat{\rho}_W\pdv{\hat{H}_W}{P})}_{R=-\infty}^{R=+\infty}+\frac{1}{2}\Tr\int\ab(\pdv{\hat{H}_W}{R,P}\hat{\rho}_W+\hat{\rho}_W\pdv{\hat{H}_W}{R,P})\odif{R,P}\nonumber\\
				&=0 ,
			\end{align}
			since the density vanishes at infinity. This property ensures that the operator \(\iu\hat{\mathcal{L}}\) is anti-Hermitian with respect to the combination of the trace and integration over phase-space coordinates.

		\subsection{Energy}
			To demonstrate the conservation of energy, it is simplest to use the anti-Hermitian properties of the Liouville operator,
			\begin{equation}
				\Tr\int\hat{A}_W(R,P)\iu\hat{\mathcal{L}}\hat{B}_W(R,P)\odif{R,P}=-\Tr\int\ab(\iu\hat{\mathcal{L}}\hat{A}_W(R,P))\hat{B}_W(R,P) \odif{R,P} ,
			\end{equation}
			which holds provided the boundary terms vanish, as demonstrated above.

			Since \(\iu\hat{\mathcal{L}}\hat{H}_W=0\), it follows that
			\begin{align}
				\pdv{\ev{\hat{H}_W}}{t}&=-\Tr\int\hat{H}_W\iu\hat{\mathcal{L}}\hat{\rho}_W(t)\odif{R,P}\\
				&=\Tr\int\ab(\iu\hat{\mathcal{L}}\hat{H}_W)\hat{\rho}_W(t)\odif{R,P}=0 .
			\end{align}

		\subsection{Purity}
			Following the definition of purity in Eq.~\ref{eq:def purity}, and using the anti-Hermitian property of the Liouville operator, we see that
			\begin{align}
				\pdv{S}{t}&=4\pi\hbar\Tr\int\hat{\rho}_W\pdv{\hat{\rho}_W}{t}\odif{R,P}\\
				&=-4\pi\hbar\Tr\int\hat{\rho}_W(t)\iu\hat{\mathcal{L}}{\hat{\rho}}_W(t)\odif{R,P}\\
				&=4\pi\hbar\Tr\int\ab(\iu\hat{\mathcal{L}} \hat{\rho}_W(t))\hat{\rho}_W(t)\odif{R,P}=0
			\end{align}
			using the cyclic property of the trace.

	\section{\label{app:elm marg pos}Positivity of Marginal Density for Diagonal Elements}
		In both the diabatic and adiabatic bases, the partial Wigner transform, Eqs.~\ref{eqs:WT}, implies the equivalence of the positional marginal density of the diagonal elements and squared norm of the wavefunction of the corresponding state,
		\begin{subequations}
			\begin{align}
				n_i(R,t)&=\int\rho_{W}^{ii}(R,P,t)\odif{P}=|\psi_i(R,t)|^2 ,\\
				n_\alpha(R,t)&=\int\rho_{W}^{\alpha\alpha}(R,P,t)\odif{P}=|\psi_\alpha(R,t)|^2 ,
			\end{align}
		\end{subequations}
		by direct integration over momentum.

		For the momentum marginal density, since the adiabatic basis states are eigenstates of \(\hat{H}_0(\hat{r};\hat{R})\) which does not commute with \(\hat{P}\), there are no common eigenstates of \(\hat{H}_0\) and \(\hat{P}\), or momentum space adiabatic wavefunction. Consequently, the marginal of \(P\) for each adiabatic surface, \(\eta_{\alpha}(P,t)\), has no physical meaning and is not guaranteed to be positive, as in Eqs.~\ref{eqs:const model init P marg}.

		On the other hand, since the diabatic basis is independent of bath coordinates, one can always apply a Fourier transform to construct a momentum-space wavefunction for each of the diabatic states,
		\begin{align}
			\ket{\Phi(P,t)}&=\frac{1}{\sqrt{2\pi\hbar}}\int\ket{\Psi(R,t)}\e{-\iu PR/\hbar}\odif{R}\\
			&=\frac{1}{\sqrt{2\pi\hbar}}\int\sum_i\psi_i(R,t)\ket{i}\e{-\iu PR/\hbar}\odif{R}=\sum_i\phi_i(P,t)\ket{i} ,
		\end{align}
		where
		\begin{equation}
			\phi_i(P,t)=\frac{1}{\sqrt{2\pi\hbar}}\int\psi_i(R,t)\e{-\iu PR/\hbar}\odif{R} ,
		\end{equation}
		satisfying
		\begin{equation}
			\eta_i(P,t)=\int\rho_{W}^{ii}(R,P,t)\odif{R}=|\phi_i(P,t)|^2 .
		\end{equation}
		It follows that the marginals \(n_i(R,t)\) and \(\eta_i(P,t)\) of each diagonal diabatic elements must be positive definite at all times. Although the marginals of the diabatic state populations are not directly observable if there is non-zero coupling between diabatic states, the dynamics should preserve their positivity. Nevertheless, this property is violated in the QCLE dynamics. Note that these considerations also apply to the joint probability densities of position \(n(R_1,\dots,R_N,t)\) and momentum \(\eta(P_1,\dots,P_N,t)\) of systems with \(N\) spatial and momentum degrees of freedom.

	\section{\label{app:numeric method}Methodology for Numerical Calculation}
		\subsection{Introduction}
			Unless otherwise specified, all quantities in this section are given in atomic units with \(\hbar=1\). Numerical results were obtained by propagating an initial Gaussian wavepacket centered at \(R_0\) on a grid. For simulations of Tully's DAC model\cite{10.1063/1.459170}, the initial wavepacket is centered at \(R_0=-15.0\). For a given central momentum \(P_0\) of the initial wavepacket, the initial standard deviation of momentum is\cite{10.1063/1.459170} \(\sigma_P=P_0/20\) and the initial standard deviation of position is determined by the minimum uncertainty principle \(\sigma_R=1/(2\sigma_P)\).

			The spatial grid size \(\Delta R\) is the same for all surfaces, chosen to be the smaller value of either a fixed specified value \(\Delta R'\) or \(k\) grid points per de Broglie wavelength,
			\begin{equation}
				\Delta R\leqslant \frac{2\pi\hbar}{k(P_0+3\sigma_P)}=\frac{4\pi}{23k}\sigma_R .
			\end{equation}
			Unless indicated otherwise, the default values are \(\Delta R'=0.05\) and \(k=2\).

			For \(N\) spatial grid points and the same number of momentum grid points, a standard fast-Fourier transform (FFT) scheme would imply a momentum spacing \(\Delta P=2\pi\hbar/(N\Delta R)\). In that case, however, when \(P_N=P_1+\floor{N/2}\Delta P\), \(\rho_W(R,P_N)\approx\rho_W(R,P_1)\) due to the discrepancy between the factor of \(\exp(2\iu PQ/\hbar)\) in the Wigner transform and the factor of \(\exp(\iu PR/\hbar)\) in the Fourier transform. To overcome this artifact, with odd \(N\), \(\Delta P\) is set to be
			\begin{equation}
				\Delta P=\frac{2\pi\hbar}{(2N-1)\Delta R} ,
			\end{equation}
			and \(P\) ranges in \(P_0\pm(N-1)\Delta P/2\). When performing the FFT, \((N-1)/2\) leading and trailing zeros are added to the (diabatic) wavefunction in accordance with this grid spacing.

			Screens and boundaries were placed between \(\pm R_{\max}\) where
			\begin{equation}
				R_{\max}=\max\ab\{2|R_0|,\frac{23k\sigma_R}{4}\} .
			\end{equation}
			The latter condition ensures comparable resolution for position and momentum
			\begin{equation}
				\sigma_P/\Delta P\geqslant23k/(4\pi) .
			\end{equation}
			With fixed spatial grid spacing \(\Delta R\), \(R_{\max}\) is adjusted to guarantee that the number of grid points \(N\) is odd (so that the \(P\) ranges symmetrically around \(P_0\), and \(R\) ranges symmetrically around \(0\)).

			The time step \(\Delta t\) satisfies\cite{KOSLOFF198335}
			\begin{equation}
				\Delta t=\dfrac{\hbar}{V_{\max}+\dfrac{\pi^2}{2M\Delta x^2}} ,
			\end{equation}
			where \(V_{\max}\) is the maximum asymptotic energy among all surfaces in the model; in the DAC model, it is \(0.05\), and in the constant-potential model, it is the parameter \(E\).

			All simulations are terminated when the average position \(\langle R\rangle\) reaches \(|R_0|\).

		\subsection{Split-Operator Method\cite{FEIT1982412, KOSLOFF198335}}
			In the split-operator method, a Trotter expansion is employed
			\begin{subequations}
				\begin{align}
					\iu\hbar\pdv{\Psi}{t}&=(\hat{T}+\hat{V})\Psi ,\\
					\Rightarrow\Psi(t+\Delta t)&=\e{-\iu(\hat{T}+\hat{V})\Delta t/\hbar}\Psi(t) ,\\
					&\approx\e{-\iu\hat{V}\Delta t/(2\hbar)}\e{-\iu\hat{T}\Delta t/\hbar}\e{-\iu\hat{V}\Delta t/(2\hbar)}\Psi(t) .
				\end{align}
			\end{subequations}
			The \(\hat{V}\) propagation is performed in the adiabatic basis where \(\hat{V}\) is diagonal, and the \(\hat{T}\) propagation\textemdash{}which involves second-order derivatives\textemdash{}is carried out in the diabatic representation using the FFT and inverse-FFT routines of PyTorch\cite{NEURIPS2019bdbca288}.

		\subsection[Discrete Variable Representation]{Discrete Variable Representation\cite{10.1063/1.462100}}
			The kinetic matrix element in the diabatic DVR basis is given by
			\begin{equation}
				T^\dia_{mn}=\frac{\hbar^2}{M\Delta R^2}\begin{dcases}
					\dfrac{N^2}{6},&\text{if }m=n\\
					\dfrac{{(-1)}^{m-n}}{{(m-n)}^2},&\text{otherwise}
				\end{dcases} ,
			\end{equation}
			where \(m\) and \(n\) labels the grid pointss. Therefore, the Hamiltonian in the diabatic DVR representation is given by
			\begin{equation}
				H^\dia_{im,jn}=V_{ij}(R_m)\delta_{mn}+T^\dia_{mn} .
			\end{equation}

			This Hermitian matrix is diagonalized
			\begin{equation}
				\bm{H}^\dia=C\bm{E}C^\dagger,\qq{with}\bm{E}=\mathrm{diag}(E_1,\dots,E_N) ,
			\end{equation}
			which implies
			\begin{equation}
				\e{-\iu\bm{H}^\dia t/\hbar}=C\e{-\iu\bm{E}t/\hbar}C^\dagger .
			\end{equation}
			The adiabatic wavefunction at a given time \(t\) is calculated by transforming the initial adiabatic wavefunction to the diabatic basis \(\Psi^\dia(0)\), using the above equation to propagate the diabatic wavefunction to time \(t\) as \(\Psi^\dia(t)=\e{-\iu\bm{H}^\dia t/\hbar}\ \Psi^\dia(0)\), and then transforming the resulting diabatic wavefunction back to the adiabatic basis.

		\subsection[QCLE]{QCLE}
			The numerical QCLE results are obtained on a two-dimensional phase space grid. The ranges and grid spacings for the spatial coordinates \(R\) and momentum coordinates \(P\), and the number of grids \(N\) for \(R\) have been discussed above. The same number of grid points is used for \(P\) so that the total number of grid points for the classical phase space is \(N^2\). This grid also serves as the phase-pace screen for the partial Wigner transform of the wavefunction results.

			The QCLE propagation is carried out in the adiabatic basis. The Liouvillian \(\iu\hat{\mathcal{L}}\) is decomposed into three components,
			\begin{equation}
				-\iu\hat{\mathcal{L}}=-\iu\hat{\mathcal{L}}^Q-\iu\hat{\mathcal{L}}^R-\iu\hat{\mathcal{L}}^P .
			\end{equation}
			where the superscripts \(R\) and \(P\) represent the corresponding derivatives, and \(Q\) represents a quantum rotation. In this basis,
			\begin{subequations}
				\begin{align}
					-\iu\mathcal{L}^Q\bm{\rho}_W(R,P,t)&=-\frac{\iu}{\hbar}\ab[V_W(R)-\iu\hbar\frac{P}{M}\bm{D},\bm{\rho}_W] ,\\
					-\iu\mathcal{L}^R\bm{\rho}_W(R,P,t)&=-\frac{P}{M}\pdv{\bm{\rho}_{W}}{R} ,\\
					-\iu\mathcal{L}^P\bm{\rho}_W(R,P,t)&=-\frac{1}{2}\ab(\bm{F}_W(R)\pdv{\bm{\rho}_W}{P}+\pdv{\bm{\rho}_W}{P}\bm{F}_W(R)) .
				\end{align}
			\end{subequations}

			The propagator of the PWTDM over a timestep \(\Delta t\) is carried out using a symmetric Trotter expansion,
			\begin{equation}
				\e{-\iu\hat{\mathcal{L}}\Delta t}=\e{-\iu\hat{\mathcal{L}}^Q\Delta t/2}\e{-\iu\hat{\mathcal{L}}^R\Delta t/2}\e{-\iu\hat{\mathcal{L}}^P\Delta t}\e{-\iu\hat{\mathcal{L}}^R\Delta t/2}\e{-\iu\hat{\mathcal{L}}^Q\Delta t/2} .
			\end{equation}

			Writing
			\begin{equation}
				V'(R,P)=V_W(R)-\iu\hbar\frac{P}{M}\bm{D} ,
			\end{equation}
			which is Hermitian and can be diagonalized at each \((R,P)\), the quantum-rotation Liouvillian propagator acts grid-wise as
			\begin{equation}
				\e{-\iu\mathcal{L}^Q\Delta t}\bm{\rho}_W(R,P)=\e{-\iu V'(R,P)\Delta t/\hbar}\bm{\rho}_W(R,P)\e{\iu V'(R,P)\Delta t/\hbar} .
			\end{equation}

			The position and momentum propagator are evaluated using FFT and inverse FFT.~After the FFT, the position propagator becomes diagonal in \(P\) and is therefore trivial; the momentum propagator, however, still involves matrix multiplication over the adiabatic-state components where the matrix \(\bm{F}_W(R)\) is Hermitian and is handled similarly to the quantum-rotation propagator.

	\section{\label{app:marg EOM}Evolution Equation for a Marginal Density Under the QCLE}
		Formally exact, closed equations for the evolution of marginal densities in the adiabatic basis can be derived from the QCLE.~Although exact, these equations are nonlocal in time and involve averages of observables over an adiabatically evolved reference system. Whereas it is difficult to gain much insight from the form of these equations, we present a general procedure to obtain the dynamics of the marginal densities below.

		We start from the QCLE represented in the adiabatic basis, Eq.~\ref{eq:adia QCLE}, written in superoperator form in which the matrix elements \(\rho_W^{\alpha\beta}\) are denoted as \(\rho_W^\mu\), with \(\mu=(\alpha,\beta)\),
		\begin{align}
			\pdv{\rho_W^{\mu}}{t}&=-\sum_{\nu}\iu\mathcal{L}_{\mu\nu}\rho_W^{\nu} ,\\
			&=-\iu\mathcal{L}^\ad_{\mu}\rho_W^{\mu}-\sum_{\nu}\mathcal{J}_{\mu\nu}\rho_W^{\nu} ,
		\end{align}
		where
		\begin{align}
			\iu\mathcal{L}^\ad_{\mu}&=\iu\omega_{\mu}+\frac{P}{M}\pdv{}{R}+F_W^{\mu}\pdv{}{P} ,\\
			&=\iu\omega_{\alpha\beta}+\frac{P}{M}\pdv{}{R}+\frac{1}{2}\ab(F_W^{\alpha\alpha}+F_W^{\beta\beta})\pdv{}{P} ,
		\end{align}
		is the standard classical evolution superoperator for the adiabatic evolution of the density matrix, and
		\begin{equation}
			\mathcal{J}_{\mu\nu}=\frac{P}{M}d_{\mu\nu}+F_W^{\mu\nu}\pdv{}{P}
		\end{equation}
		is the off-diagonal transition operator between density matrix elements, where
		\begin{align}
			d_{\mu\nu}&=d_{\alpha\alpha'}\delta_{\beta\beta'}-d_{\beta'\beta}\delta_{\alpha'\alpha} ,\\
			F_W^{\mu\nu}&=\frac{1}{2}(F_W^{\alpha\alpha'}\delta_{\beta\beta'}+F_W^{\beta'\beta}\delta_{\alpha'\alpha}-(F_W^{\alpha\alpha}+F_W^{\beta\beta})\delta_{\alpha\alpha'}\delta_{\beta\beta'}) .
		\end{align}

		By direct evaluation, the evolution of marginal densities can be obtained by partial integration of the QCLE over one of the phase-space variables, for example,
		\begin{subequations}
			\begin{align}
				\pdv{n_{\mu}(R,t)}{t}&=-\sum_{\nu}\int\iu\mathcal{L}^R_{\mu\nu}\rho_W^{\nu}(R,P,t)\odif{P} ,\\
				\pdv{\eta_{\mu}(P,t)}{t}&=\sum_{\nu}\int\iu\mathcal{L}^P_{\mu\nu}\rho_W^{\nu}(R,P,t)\odif{R} ,
			\end{align}
		\end{subequations}
		where
		\begin{subequations}\label{eqs:liouvillian w superscript}
			\begin{align}
				\iu\mathcal{L}^R_{\mu\nu}&=\iu\omega_{\mu}\delta_{\mu\nu}+\frac{1}{M}\ab(\delta_{\mu\nu}\pdv{}{R}+d_{\mu\nu})P ,\\
				\iu\mathcal{L}^P_{\mu\nu}&=\iu\omega_{\mu}\delta_{\mu\nu}+\frac{P}{M}d_{\mu\nu}+(F_W^{\mu\nu}+F_W^{\mu}\delta_{\mu\nu})\pdv{}{P} .
			\end{align}
		\end{subequations}
		These equations are not closed for the marginal densities. To construct closed equations, we introduce a projection method. In what follows, we derive evolution equations for the marginal adiabatic elements \(\eta_\mu(P,t)\). A similar prescription follows to derive closed equations for the position marginals \(n_\mu(R,t)\).

		We start by defining the density matrix element for an adiabatically evolved system,
		\begin{equation}
			\rho_{W,\ad}^{\mu}(R,P,t)=\e{-\iu\mathcal{L}_\ad^{\mu}t}\rho_W^{\mu}(R,P,0) ,
		\end{equation}
		where the evolution of element \(\mu\) is \textit{entirely} adiabatic,
		and the bath normalized total marginal densities under the adiabatic evolution are,
		\begin{subequations}
			\begin{align}
				n^{\ad}(R,t)&=\sum_{\alpha}\int\rho_{W,\ad}^{\alpha\alpha}(R,P,t)\odif{P} ,\\
				\eta^{\ad}(P,t)&=\sum_{\alpha}\int\rho_{W,\ad}^{\alpha\alpha}(R,P,t)\odif{R} .
			\end{align}
		\end{subequations}

		We also define a time-dependent projection operator \(\mathcal{P}(\tau)\),
		\begin{equation}
			\mathcal{P}(\tau)f(R,P,t)=n^{\ad}(R,\tau)\int f(R,P,t)\odif{R} ,
		\end{equation}
		so that
		\begin{equation}
			\mathcal{P}(\tau)\rho_W^{\mu}(R,P,t)=n^{\ad}(R,\tau)\eta_{\mu}(P,t) .
		\end{equation}
		We note that other projection operators, such as those that utilize the stationary density of the Liouville superoperator, can be defined if the system decoheres or equilibrates quickly in either the position or momentum space\cite{Grunwald2009, 10.1063/1.2567164}.

		Since the adiabatic marginal density \(n_{\ad}(R,t)\) is normalized at all times, we have
		\begin{equation}
			\mathcal{P}(\tau)\mathcal{P}(\tau')=\mathcal{P}(\tau) ,
		\end{equation}
		which indicates that \(\mathcal{P}\) is an adiabatic projection superoperator. Its complementary projector is denoted as \(\mathcal{Q}(t)=1-\mathcal{P}(t)\), which satisfies \({\mathcal P}(\tau){\mathcal Q}(\tau')=0\) for all times \(\tau'\).

		The time dependence of the projector is given by
		\begin{align}
			\pdv{\mathcal{P}(\tau)}{\tau}f(R,P,t)&=\pdv{n^{\ad}(R,\tau)}{\tau}\int f(R,P,t)\odif{R}\nonumber\\
			&=-\pdv{v^{\ad}(R,\tau)}{R}\int f(R,P,t)\odif{R} ,
		\end{align}
		where
		\begin{equation}
			v^{\ad}(R,t)=\sum_\alpha\int\frac{P}{M}\rho^\alpha_{W,a}(R,P,t)\odif{P}
		\end{equation}
		is the average adiabatic velocity at time \(t\). Note that the projector satisfies \(\pdv{\mathcal{P}(t)}/{t}=\mathcal{Q}(t)\pdv{\mathcal{P}(t)}/{t}\mathcal{P}(t)\) and hence \(\ab(\pdv{\mathcal{P}(t)}/{t})\mathcal{Q}(t)=0\).

		Denoting \(Z^{\mu}(R,P,t)=\mathcal{Q}(t)\rho_W^{\mu}(R,P,t)\) and separating the projections of \(\rho_W^\mu\), we obtain
		\begin{equation}
			\rho_W^{\mu}(R,P,t)=n^{\ad}(R,t)\eta_{\mu}(P,t)+Z^{\mu}(R,P,t) .
		\end{equation}

		The evolution of \(Z^\mu\) is given by
		\begin{align}
			\pdv{Z^{\mu}}{t}&=\pdv*{\ab[(1-\mathcal{P}(t))\rho_W^{\mu}(t)]}{t}\nonumber\\
			&=\ab(\mathcal{Q}(t)\pdv{}{t}-\pdv{\mathcal{P}(t)}{t})(\mathcal{P}(t)+\mathcal{Q} (t))\rho_W^{\mu}(t)\nonumber\\
			&=-\sum_{\nu}\mathcal{Q}(t)\iu\mathcal{L}_{\mu\nu}Z^{\nu}-\pdv{\mathcal{P}(t)}{t}(1-\mathcal{P}(t))\rho_W^{\mu}-\mathcal{Q}(t)\sum_{\nu}\ab(\iu\mathcal{L}_{\mu\nu}+\pdv{\mathcal{P}(t)}{t}\delta_{\mu\nu})\mathcal{P}(t) \rho_W^{\nu}\nonumber\\
			&=-\sum_{\nu}\mathcal{Q}(t)\iu\mathcal{L}_{\mu\nu}Z^{\nu}(R,P,t)-\mathcal{Q}(t)\sum_{\nu}\ab(\iu\mathcal{L}_{\mu\nu}+\pdv{\mathcal{P}(t)}{t}\delta_{\mu\nu})n^{\ad}(R,t)\eta_{\nu}(P,t) .
		\end{align}

		Defining the projected evolution operator
		\begin{equation}
			\mathcal{U}^{\mu\nu}(t,\tau)={\mathcal{T}_+\exp\ab(-\int_{\tau}^t\odif{t'}\mathcal{Q}(t')\iu\mathcal{L})}_{\mu\nu}\label{eq:Dyson exp} ,
		\end{equation}
		where \(\mathcal{T}_+\) is the time-ordering operator satisfying
		\begin{equation}
			\mathcal{T}_+A(t_1)A(t_2)=\begin{cases}
				A(t_1)A(t_2),\qif t_1>t_2\\
				A(t_2)A(t_1),\qif t_1\leqslant t_2 ,
			\end{cases}
		\end{equation}
		we obtain
		\begin{multline}
			Z^{\mu}(R,P,t)=\sum_{\nu}\left(\mathcal{U}^{\mu\nu}(t,0)Z^{\nu}(R,P,0)-\int_{0}^{t}\odif{\tau}\ \mathcal{U}^{\mu\nu}(t,\tau)\right.\\
			\left.\times\mathcal{Q}(\tau)\sum_{\nu'}\ab(\iu\mathcal{L}_{\nu\nu'}+\pdv{\mathcal{P} (\tau)}{\tau}\delta_{\nu\nu'})n^{\ad}(R,\tau)\eta_{\nu'}(P,\tau)\right)\label{eq:Z expression} ,
		\end{multline}
		where
		\begin{equation}
			Z^{\mu}(R,P,0)=(1-\mathcal{P}(0))\rho_W^{\mu}(R,P,0)=\rho_W^{\mu}(R,P,0)-n^{\ad}(R,0)\eta_\mu(P,0) ,
		\end{equation}
		which vanishes if the system is initially in the ground adiabatic state with negligible coupling and if position and momentum are independently distributed. When the initial value term \(Z^\nu (R,P,0)\) vanishes, the evolution of \(Z^\mu\) is solely determined by an operator equation for the marginal density \(\eta(P,t')\). This property enables a closed equation for the evolution of \(\eta_{\mu}(P,t)\) to be obtained.

		The evolution of the marginal density \(\eta_\mu(P,t)\) is given by,
		\begin{align}
			\pdv{\eta_{\mu}(P,t)}{t}&=-\int\sum_{\nu}\iu\mathcal{L}_{\mu\nu}\rho_W^{\nu}\odif{R} ,\\
			&=-\sum_{\nu}\int\iu\mathcal{L}^P_{\mu\nu}\ab[n^{\ad}(R,t)\eta_{\nu}(P,t)+Z^{\nu}(R,P,t)]\odif{R} .
		\end{align}
		Substituting Eq.~\ref{eq:Z expression} into the above equation and neglecting the initial value term \(Z^{\mu}(R,P,0)\), we obtain a closed evolution equation for the marginal densities \(\eta_\mu(P,t)\),
		\begin{multline}
			\pdv{\eta_{\mu}(P,t)}{t}=-\sum_{\nu}\ev{\iu\mathcal{L}^P_{\mu\nu}}_{t}\eta_{\nu}(P,t)\\
			+\int_{0}^{t}\odif{\tau}\sum_{\nu,\nu',\nu''}\ev{\iu\mathcal{L}^P_{\mu\nu}\mathcal{U}^{\nu\nu'}(t,\tau)\mathcal{Q}(\tau)\ab(\iu\mathcal{L}_{\nu'\nu''}+\pdv{\mathcal{P}(\tau)}{\tau}\delta_{\nu',\nu''})}_{\tau}\eta_{\nu''}(P,\tau) ,
		\end{multline}
		where \(\langle\hat{O}(R,P)\rangle_t=\int\hat{O}(R,P)n^{\ad}(R,t)\odif{R}\).

		Similarly, using the projector
		\begin{equation}
			\tilde{\mathcal{P}}(t)f(R,P,t)=\eta^{\ad}(P,t)\int f(R,P,t)\odif{P}
		\end{equation}
		and its complement \(\tilde{\mathcal{Q}}(t)=1-\tilde{\mathcal{P}}(t)\), the marginal density \(n_\mu(R,t)\) obeys
		\begin{multline}
			\pdv{n_{\mu}(R,t)}{t}=-\sum_{\nu}\ev{\iu\mathcal{L}^R_{\mu\nu}}_{t}n_{\nu}(R,t)\\
			+\int_{0}^{t}\odif{\tau}\sum_{\nu,\nu',\nu''}\ev{\iu\mathcal{L}^R_{\mu\nu}\tilde{\mathcal{U}}^{\nu\nu'}(t,\tau)\tilde{\mathcal{Q}}(\tau)\ab(\iu\mathcal{L}_{\nu'\nu''}+\pdv{\tilde{\mathcal{P}}(\tau)}{\tau}\delta_{\nu',\nu''})}_{\tau}n_{\nu''}(R,\tau)\nonumber ,
		\end{multline}
		where now \(\langle\hat{O}(R,P)\rangle_t=\int\hat{O}(R,P)\eta^{\ad}(P,t)\odif{P}\) and the projected evolution operator \(\tilde{U}(t,\tau)\) is defined as in Eq.~\ref{eq:Dyson exp} with \(\mathcal{Q}(t')\) replaced by \(\tilde{\mathcal{Q}}(t')\).

		Note that these evolution equations for the marginal densities depend on knowledge of the diagonal elements of the adiabatically evolved density. Other choices of projection operators may be more physically useful. For example, if the marginals of the classical momentum rapidly approach a stationary form, a projection operator can be defined relative to marginals of the stationary solution \(\iu\hat{\mathcal{L}}\hat{\rho}_W^e(R,P)=0\) so that
		\begin{align}
			\eta^e(P)&=\int\Tr\hat{\rho}_{W,e}(R,P)\odif{R}\nonumber\\
			\tilde{\mathcal{P}^e}f(R,P)&=\eta^e(P)\int f(R,P)\odif{P} ,
		\end{align}
		then \(\pdv{\tilde{\mathcal{P}^e}}/{\tau}=0\) and coupled Smoluchowski-type equations\cite{SCHOFIELD1993209} are obtained for the marginals \(n_\mu(R,t)\).

	\section{Analysis of the Constant Model}
		\subsection{\label{app:const TDSE sol}Solution of \sch{} Equation for the Constant Model}
			The time-dependent \sch{} equation represented in an adiabatic basis,
			\begin{multline}
				\iu\hbar\pdv{\psi_\alpha(R,t)}{t}=E_\alpha\psi_\alpha(R,t)-\frac{\hbar^2}{2M}\sum_\beta g_{\alpha\beta}(R)\psi_\beta(R,t)\\
				-\frac{\hbar^2}{M}\sum_\beta d_{\alpha\beta}(R)\pdv{\psi_\beta(R,t)}{R}-\frac{\hbar^2}{2M}\pdv[]{\psi_\alpha(R,t)}{R}
			\end{multline}
			gives
			\begin{subequations}
				\begin{align}
					\iu\hbar\pdv{\psi_0(R,t)}{t}&=\ab(\frac{\hbar^2 D^2}{2M}-\frac{\hbar^2}{2M}\pdv[2]{}{R})\psi_0(R,t)-\frac{\hbar^2 D}{M}\pdv*{\psi_1(R,t)}{R}\\
					\iu\hbar\pdv{\psi_1(R,t)}{t}&=\ab(E+\frac{\hbar^2 D^2}{2M}-\frac{\hbar^2}{2M}\pdv[2]{}{R})\psi_1(R,t)+\frac{\hbar^2 D}{M}\pdv*{\psi_0(R,t)}{R}
				\end{align}
			\end{subequations}
			for a constant nonadiabatic coupling matrix element \(d_{12}(R)=D\), with the initial condition in Eq.~\ref{eq:const init wfn} with \(\theta=\pi/4\).

			Adopting the Fourier transform,
			\begin{equation}
				\Phi(\xi,t)=\frac{1}{\sqrt{2\pi\hbar}}\int\Psi(R,t)\e{-\iu\xi R/\hbar}\odif{R} ,
			\end{equation}
			we have
			\begin{subequations}\label{eqs:se constant}
				\begin{align}
					\iu\hbar\pdv{\phi_0(\xi,t)}{t}&=\ab(\frac{\hbar^2 D^2}{2M}+\frac{\xi^2}{2M})\phi_0(\xi,t)-\frac{\iu\hbar D\xi}{M}\phi_1(\xi,t)\\
					\iu\hbar\pdv{\phi_1(\xi,t)}{t}&=\frac{\iu\hbar D\xi}{M}\phi_0(\xi,t)+\ab(E+\frac{\hbar^2 D^2}{2M}+\frac{\xi^2}{2M})\phi_1(\xi,t) .
				\end{align}
			\end{subequations}
			Notice that although \(\xi\) and \(P\) are of the same dimension, \(\phi(\xi)\) is not the adiabatic wavefunction in the momentum representation but rather is the Fourier transform of the adiabatic spatial wavefunction, as explained in Appendix~\ref{app:elm marg pos}.

			The solution of Eqs.~\ref{eqs:se constant} is given by
			\begin{multline}\label{eq:const model exact TDSE solution}
				\begin{pmatrix}\phi_0(\xi,t)\\\phi_1(\xi,t)\end{pmatrix}=\frac{1}{\zeta}\exp\ab(-\frac{\iu t}{2M}(\xi^2+\hbar^2 D^2+E^2 M^2))\\
				\times\begin{pmatrix}
					\zeta\cos\tau+\iu EM\sin\tau & -2\hbar D\xi\sin\tau\\
					2\hbar D\xi\sin\tau & \zeta\cos\tau-\iu EM\sin\tau
				\end{pmatrix}\begin{pmatrix}\phi_0(\xi,0)\\\phi_1(\xi,0)\end{pmatrix} ,
			\end{multline}
			where
			\begin{equation}
				\zeta=\sqrt{4\hbar^2 D^2 \xi^2+E^2 M^2}
			\end{equation}
			and \(\tau=\zeta t/(2M)\).

			The Fourier-transformed initial conditions are
			\begin{equation}
				\begin{pmatrix}\phi_0(\xi,0)\\\phi_1(\xi,0)\end{pmatrix}={(2\pi)}^{-\frac{1}{4}}\sigma_P^{-\frac{1}{2}}\exp\ab(-{\ab(\frac{\xi-P_0}{2\sigma_P})}^2-\frac{\iu R_0(\xi-P_0)}{\hbar})\begin{pmatrix}\dfrac{1}{\sqrt{2}}\\\dfrac{1}{\sqrt{2}}\end{pmatrix} .
			\end{equation}

			The spatial wavefunctions \(\Psi(R,t)\) cannot be evaluated explicitly, but can be computed numerically using an inverse Fourier transform,
			\begin{equation}
				\Psi(R,t)=\frac{1}{\sqrt{2\pi\hbar}}\int\Phi(\xi,t)\e{\iu R\xi/\hbar}\odif{\xi} .
			\end{equation}
			Using these solutions, the adiabatic PWTDM can be constructed following the procedure in Eqs.~\ref{eqs:WT}.

		\subsection{Dimensionless Form of the QCLE}
			We first introduce a dimensionless form of the QCLE to facilitate perturbative analysis. As \(E\) is a constant in the model, we define \(\tilde{t}= Et/\hbar\) to be the dimensionless time. For position and momentum, the initial standard deviation is used for nondimensionalization,
			\begin{subequations}
				\begin{align}
					\tilde{R}=\frac{R}{\sigma_R}\\
					\tilde{P}=\frac{P}{\sigma_P} ,
				\end{align}
			\end{subequations}
			where the standard deviations follow the minimum uncertainty rule \(\sigma_R\sigma_P=\hbar/2\).

			We further define two constants,
			\begin{subequations}
				\begin{align}
					c_1&=\frac{\hbar D}{\sigma_P}=2D\sigma_R\\
					c_2&=\frac{\sigma_P^2}{EM} ,
				\end{align}
			\end{subequations}
			so that Eqs.~\ref{eqs:const model qcle}, the QCLE of this model, is written in dimensionless variables as,
			\begin{subequations}
				\begin{align}
					\pdv{\rho_W^{00}}{\tilde{t}}&=\frac{c_1}{2}\pdv{(\rho_W^{01}+\rho_W^{10})}{\tilde{P}}-c_1 c_2\tilde{P}\ab(\rho_W^{01}+\rho_W^{10})-2c_2\tilde{P}\pdv{\rho_W^{00}}{\tilde{R}}\\
					\pdv{\rho_W^{10}}{\tilde{t}}&=\frac{c_1}{2}\pdv{(\rho_W^{00}+\rho_W^{11})}{\tilde{P}}+c_1 c_2\tilde{P}\ab(\rho_W^{00}-\rho_W^{11})-\ab(\iu+2c_2\tilde{P}\pdv{}{\tilde{R}})\rho_W^{10}\\
					\pdv{\rho_W^{11}}{\tilde{t}}&=\frac{c_1}{2}\pdv{(\rho_W^{01}+\rho_W^{10})}{\tilde{P}}+c_1 c_2\tilde{P}\ab(\rho_W^{01}+\rho_W^{10})-2c_2\tilde{P}\pdv{\rho_W^{11}}{\tilde{R}} ,
				\end{align}
			\end{subequations}
			and Eqs.~\ref{eqs:const model qcle p marg evo} become
			\begin{subequations}\label{eqs:const model dimless qcle p marg}
				\begin{align}
					\pdv{\eta_0}{\tilde{t}}&=c_1\ab(\pdv{}{\tilde{P}}-2c_2\tilde{P})\eta_r\\
					\pdv{\eta_r}{\tilde{t}}&=\frac{c_1}{2}\ab(\ab(\pdv{}{\tilde{P}}+2c_2\tilde{P})\eta_0+\ab(\pdv{}{\tilde{P}}-2c_2\tilde{P})\eta_1)+\eta_i\\
					\pdv{\eta_i}{\tilde{t}}&=-\eta_r\\
					\pdv{\eta_1}{\tilde{t}}&=c_1\ab(\pdv{}{\tilde{P}}+2c_2\tilde{P})\eta_r .
				\end{align}
			\end{subequations}
			In supermatrix form, Eqs.~\ref{eqs:const model dimless qcle p marg} is denoted as as \(\pdv{\bm{\eta}}/{\tilde{t}}=\mathcal{L}_P\bm{\eta}\) where \(\bm{\eta}=\begin{pmatrix}\eta_0&\eta_r&\eta_i&\eta_1\end{pmatrix}^{\mathsf{T}}\) and
			\begin{equation}
				\mathcal{L}_P=\begin{pmatrix}
					0&\tilde{\mathcal{D}}_-&0&0\\
					\tilde{\mathcal{D}}_+/2&0&1&\tilde{\mathcal{D}}_-/2\\
					0&-1&0&0\\
					0&\tilde{\mathcal{D}}_+&0&0
				\end{pmatrix} ,
			\end{equation}
			with \(\tilde{\mathcal{D}}_\pm=c_1(\pdv{}/{\tilde{P}}\pm2c_2\tilde{P})\). Based on the nature of the operators, \(\mathcal{L}_P\) can be separated into \(\mathcal{L}_P=\LP{diff}+\LP{rot}+\LP{mul}\) where
			\begin{align}
				\LP{diff}&=c_1\LPM{diff}\pdv{}{\tilde{P}}=\frac{c_1}{2}\begin{pmatrix}
					0&2&0&0\\
					1&0&0&1\\
					0&0&0&0\\
					0&2&0&0
				\end{pmatrix}\pdv{}{\tilde{P}}\label{eq:mdiff}\\
				\LP{rot}&=\LPM{rot}=\begin{pmatrix}
					0&0&0&0\\
					0&0&1&0\\
					0&-1&0&0\\
					0&0&0&0
				\end{pmatrix}\label{eq:mrot}\\
				\LP{mul}&=2c_1 c_2\tilde{P}\LPM{mul}=c_1 c_2\tilde{P}\begin{pmatrix}
					0&-2&0&0\\
					1&0&0&-1\\
					0&0&0&0\\
					0&2&0&0
				\end{pmatrix}\label{eq:mmult} .
			\end{align}

		\subsection{\label{app:const model QCLE vs QLE}Comparison of the QCLE and Liouville-von Neumann Equations}
			\newcommand{\pdrad}[2]{{\ab(\pdv[#1]{#2}{R})}}
			Starting from the adiabatic Hamiltonian, we have
			\begin{align}
				\pdv{\hat{H}_W}{P}&=\frac{P}{M}I\\
				\pdv[2]{\hat{H}_W}{P}&=\frac{1}{M}I\\
				\pdv[3]{\hat{H}_W}{P}&=\hat{0} ,
			\end{align}
			and
			\begin{align}
				\pdv{\braket[3]{\alpha}{\pdv[n]{\hat{H}_W}/{R}}{\beta}}{R}&=\sum_{\gamma}\pdv{\bra{\alpha}}{R}\ket{\gamma}\braket[3]{\gamma}{\pdv[n]{\hat{H}_W}{R}}{\beta}\nonumber\\
				&+\braket[3]{\alpha}{\pdv[n+1]{\hat{H}_W}{R}}{\beta}+\sum_{\gamma}\braket[3]{\alpha}{\pdv[n]{\hat{H}_W}{R}}{\gamma}\bra{\gamma}\pdv{\ket{\beta}}{R}\\
				\Rightarrow\pdrad{n+1}{H}&=\pdv*{\pdrad{n}{H}}{R}+\bm{D}\pdrad{n}{H}-\pdrad{n}{H}\bm{D}\\
				&=\pdv*{\pdrad{n}{H}}{R}+\ab[\bm{D},\pdrad{n}{H}] ,
			\end{align}
			where \([\cdot,\cdot]\) is the commutator, and
			\begin{align}
				\bm{D}&=D\begin{pmatrix}0&1\\-1&0\end{pmatrix}=\iu D\sigma_2\\
				H_W&=E\begin{pmatrix}0&0\\0&1\end{pmatrix}=\frac{E}{2}(I-\sigma_3) ,
			\end{align}
			so that \(\forall n\in\mathbb{Z}^+\),
			\begin{equation}
				\pdrad{n}{H}={(-1)}^{\floor{(n-1)/2}}2^{n-1}D^n E\begin{cases}
					\sigma_1,\qif n\equiv 1\pmod2\\
					\sigma_3,\qif n\equiv 0\pmod2
				\end{cases} ,
			\end{equation}
			where \(\sigma_i\) are the Pauli matrices.

			Using these results, the partial Wigner transform of Liouville-von Neumann equation in the adiabatic basis is
			\begin{align}
				\pdv{\rho_W}{t}&=-\frac{\iu}{\hbar}\ab(H_W\exp\ab(\frac{\hbar\Lambda}{2\iu})\rho_W-\rho_W\exp\ab(\frac{\hbar\Lambda}{2\iu})H_W)\\
				&=-\frac{\iu}{\hbar}(H_W\rho_W-\rho_W H_W)-\frac{\iu}{\hbar}\frac{\hbar}{2\iu}\ab(\pdv{H_W}{P}\pdv{\rho_W}{R}-\pdv{H}{R}\pdv{\rho_W}{P}-\pdv{\rho_W}{P}\pdv{H}{R}+\pdv{\rho_W}{R}\pdv{H_W}{P})\nonumber\\
				&\quad-\frac{\iu}{\hbar}\sum_{n=2}^{+\infty}\frac{1}{n!}{\ab(\frac{\hbar}{2\iu})}^n\!\ab({(-1)}^{n}\pdv[n]{H}{R}\pdv[n]{\rho_W}{P}-\pdv[n]{\rho_W}{P}\pdv[n]{H}{R}) .
			\end{align}
			The first two terms constitute the QCLE, while the rest are higher-order corrections to the QCLE.\@

			In dimensionless form, this simplifies to
			\begin{multline}
				\pdv{\rho_W}{\tilde{t}}=-2c_2\tilde{P}\pdv{\rho_W}{\tilde{R}}-\iu c_1 c_2\tilde{P}[\sigma_2,\rho_W]\\
				+\frac{1}{2}\sum_{n=0}^{+\infty}\frac{c_1^{2n+1}}{(2n+1)!}\pdv*[2n+1]{{\{\sigma_1,\rho_W\}}_+}{\tilde{P}}+\iu\sum_{n=0}^{+\infty}\frac{c_1^{2n}}{(2n)!}\pdv[2n]{}{\tilde{P}}\frac{[\sigma_3,\rho_W]}{2} ,
			\end{multline}
			where \({\{A,B\}}_+=AB+BA\) and
			\begin{subequations}
				\begin{align}
					{\ab\{\sigma_1,A\}}_+&=\begin{pmatrix}A_{01}+A_{10}&A_{00}+A_{11}\\A_{00}+A_{11}&A_{01}+A_{10}\end{pmatrix}\\
					-\iu\ab[\sigma_2,A]&=\begin{pmatrix}-(A_{01}+A_{10})&A_{00}-A_{11}\\A_{00}-A_{11}&A_{01}+A_{10}\end{pmatrix}\\
					\ab[\sigma_3, A]&=2\begin{pmatrix}0&A_{01}\\-A_{10}&0\end{pmatrix} .
				\end{align}
			\end{subequations}

			Inserting the above equations into \(\pdv{\rho_W}/{\tilde{t}}\), gives
			\begin{multline}
				\pdv{}{\tilde{t}}\begin{pmatrix}\rho_W^{00}\\\rho_W^{01}\\\rho_W^{10}\\\rho_W^{11}\end{pmatrix}=-2c_2\tilde{P}\pdv{}{\tilde{R}}\begin{pmatrix}\rho_W^{00}\\\rho_W^{01}\\\rho_W^{10}\\\rho_W^{11}\end{pmatrix}+c_1 c_2\tilde{P}\begin{pmatrix}-(\rho_W^{01}+\rho_W^{10})\\\rho_W^{00}-\rho_W^{11}\\\rho_W^{00}-\rho_W^{11}\\\rho_W^{01}+\rho_W^{10}\end{pmatrix}\\
				+\frac{1}{2}\sum_{n=0}^{+\infty}\frac{c_1^{2n+1}}{(2n+1)!}\pdv[2n+1]{}{\tilde{P}}\begin{pmatrix}\rho_W^{01}+\rho_W^{10}\\\rho_W^{00}+\rho_W^{11}\\\rho_W^{00}+\rho_W^{11}\\\rho_W^{01}+\rho_W^{10}\end{pmatrix}+\iu\sum_{n=0}^{+\infty}\frac{c_1^{2n}}{(2n)!}\pdv[2n]{}{\tilde{P}}\begin{pmatrix}0\\\rho_W^{01}\\-\rho_W^{10}\\0\end{pmatrix} .
			\end{multline}

			Since
			\begin{subequations}
				\begin{align}
					\sum_{n=0}^{+\infty}\frac{c^{2n+1}}{(2n+1)!}\pdv[2n+1]{f}{X}&=\sinh\ab(c\pdv{}{X})f(X)\nonumber\\
					&=\frac{\e{c\pdv{}/{X}}-\e{-c\pdv{}/{X}}}{2}f(X)=\frac{f(X+c)-f(X-c)}{2}\\
					\sum_{n=0}^{+\infty}\frac{c^{2n}}{(2n)!}\pdv[2n]{f}{X}&=\cosh\ab(c\pdv{}{X})f(X)\nonumber\\
					&=\frac{\e{c\pdv{}/{X}}+\e{-c\pdv{}/{X}}}{2}f(X)=\frac{f(X+c)+f(X-c)}{2} ,
				\end{align}
			\end{subequations}
			we have \(\rho_W^{01}={(\rho_W^{10})}^*\), and
			\begin{subequations}
				\begin{align}
					\pdv{\rho_W^{00}(\tilde{R},\tilde{P},\tilde{t})}{\tilde{t}}&=-2c_2\tilde{P}\pdv{\rho_W^{00}(\tilde{R},\tilde{P},\tilde{t})}{\tilde{R}}-c_1 c_2\tilde{P}\bigg(\rho_W^{01}(\tilde{R},\tilde{P},\tilde{t})+\rho_W^{10}(\tilde{R},\tilde{P},\tilde{t})\bigg)\nonumber\\
					&+\frac{1}{4}\bigg(\rho_W^{01}(\tilde{R},\tilde{P}+c_1,\tilde{t})+\rho_W^{10}(\tilde{R},\tilde{P}+c_1,\tilde{t})-\rho_W^{01}(\tilde{R},\tilde{P}-c_1,\tilde{t})-\rho_W^{10}(\tilde{R},\tilde{P}-c_1,\tilde{t})\bigg)\\
					\pdv{\rho_W^{10}(\tilde{R},\tilde{P},\tilde{t})}{\tilde{t}}&=-\frac{\iu}{2}\bigg(\rho_W^{10}(\tilde{R},\tilde{P}+c_1,\tilde{t})+\rho_W^{10}(\tilde{R},\tilde{P}-c_1,\tilde{t})\bigg)\nonumber\\
					&-2c_2\tilde{P}\pdv{\rho_W^{10}(\tilde{R},\tilde{P},\tilde{t})}{\tilde{R}}+c_1 c_2\tilde{P}\bigg(\rho_W^{00}(\tilde{R},\tilde{P},\tilde{t})-\rho_W^{11}(\tilde{R},\tilde{P},\tilde{t})\bigg)\nonumber\\
					&+\frac{1}{4}\bigg(\rho_W^{00}(\tilde{R},\tilde{P}+c_1,\tilde{t})+\rho_W^{11}(\tilde{R},\tilde{P}+c_1,\tilde{t})-\rho_W^{00}(\tilde{R},\tilde{P}-c_1,\tilde{t})-\rho_W^{11}(\tilde{R},\tilde{P}-c_1,\tilde{t})\bigg)\\
					\pdv{\rho_W^{11}(\tilde{R},\tilde{P},\tilde{t})}{\tilde{t}}&=-2c_2\tilde{P}\pdv{\rho_W^{11}(\tilde{R},\tilde{P},\tilde{t})}{\tilde{R}}+c_1 c_2\tilde{P}\bigg(\rho_W^{01}(\tilde{R},\tilde{P},\tilde{t})+\rho_W^{10}(\tilde{R},\tilde{P},\tilde{t})\bigg)\nonumber\\
					&+\frac{1}{4}\bigg(\rho_W^{01}(\tilde{R},\tilde{P}+c_1,\tilde{t})+\rho_W^{10}(\tilde{R},\tilde{P}+c_1,\tilde{t})-\rho_W^{01}(\tilde{R},\tilde{P}-c_1,\tilde{t})-\rho_W^{10}(\tilde{R},\tilde{P}-c_1,\tilde{t})\bigg) .
				\end{align}
			\end{subequations}
			These results appear in the main text in unscaled coordinates as Eq.~\ref{eqs:const model exact}.

		\subsection{\label{app:const pert sol}Perturbative Solution of Momentum Marginal Density from QCLE}
			In this section, we outline the derivation of the perturbative results for the marginal momentum densities discussed in Sec.~\ref{sec:const model}. To analyze the evolution equations of the marginal density, we compare the magnitude of \(c_1=\hbar D/\sigma_P\), \(\e{}\), and \(c_1 c_2|\tilde{P}_0|=\hbar D |P_0|/(EM)\) that scale the operators \(\LP{diff}\), \(\LP{rot}\), and \(\LP{mul}\), respectively, to distinguish which terms are dominant, denoted by \(\LP{d}\), and which others are perturbative, \(\LP{p}\). The marginal density is then written in a perturbation series,
			\begin{equation}
				\bm{\eta}(\tilde{P},\tilde{t})=\sum_{m}\bm{\eta}^{(m)}(\tilde{P},\tilde{t})=\bm{\eta}^{(0)}(\tilde{P},\tilde{t})+\bm{\eta}^{(1)}(\tilde{P},\tilde{t})+\dots ,
			\end{equation}
			where
			\begin{subequations}
				\begin{align}
					\pdv{\bm{\eta}^{(0)}}{\tilde{t}}&=\LP{d}\bm{\eta}^{(0)}\\
					\pdv{\bm{\eta}^{(m)}}{\tilde{t}}&=\LP{d}\bm{\eta}^{(m)}+\LP{p}\bm{\eta}^{(m-1)}\quad\forall m\in\mathbb{Z}^+ ,
				\end{align}
			\end{subequations}
			with initial conditions,
			\begin{subequations}
				\begin{align}
					\bm{\eta}^{(0)}(\tilde{P},0) &= \; \bm{\eta}(\tilde{P},0)\\
					\bm{\eta}^{(m)}(\tilde{P},0) &= \; 0 .
				\end{align}
			\end{subequations}
			Formally, the solution of the perturbation series is
			\begin{subequations}
				\begin{align}
					\bm{\eta}^{(0)}(P,t)&=\e{\LP{d}t}\bm{\eta}(P,0)\\
					\bm{\eta}^{(m)}(P,t)&=\int_0^{t}\odif{\tau}\e{\LP{d}(t-\tau)}\LP{p}\bm{\eta}^{(m-1)}(P,\tau) .
				\end{align}
			\end{subequations}

			When \(c_1\) is large, \(\LP{d}=\LP{diff}\) is the principal term, and the zeroth-order solution is
			\begin{equation}
				\bm{\eta}(\tilde{P},\tilde{t})=\mathcal{M}_+\bm{\eta}(\tilde{P}+c_1\tilde{t},0)+\mathcal{M}_0\bm{\eta}(\tilde{P},0)+\mathcal{M}_-\bm{\eta}(\tilde{P}-c_1\tilde{t},0) ,
			\end{equation}
			where
			\begin{subequations}
				\begin{align}
					\mathcal{M}_\pm&=\mathcal{M}_\pm^\leftarrow\mathcal{M}_\pm^\rightarrow\\
					\mathcal{M}_\pm^\leftarrow&=\frac{1}{2}\begin{pmatrix}1&\pm1&0&1\end{pmatrix}^\mathsf{T}\\
					\mathcal{M}_\pm^\rightarrow&=\frac{1}{2}\begin{pmatrix}1&\pm2&0&1\end{pmatrix}\\
					\mathcal{M}_0&=\frac{1}{2}\begin{pmatrix}1&0&0&-1\\0&0&0&0\\0&0&2&0\\-1&0&0&1\end{pmatrix} .
				\end{align}
			\end{subequations}
			Notice that the initial condition can be expressed as a combination of Gaussian wavepackets
			\begin{equation}
				\bm{\eta}(\tilde{P},0)=\sum_i\bm{w}_i N_0({\tilde{P}-\tilde{P}_i})
			\end{equation}
			where we denote
			\begin{subequations}
				\begin{align}
					N_k(x)&=x^k\exp\ab(-\frac{x^2}{2})\\
					M_k(x)&=x^k\erf\ab(\frac{x}{\sqrt{2}}) ,
				\end{align}
			\end{subequations}
			so that the first-order correction gives
			\begin{align}
				\bm{\eta}^{(1)}(\tilde{P},\tilde{t})&=\sum_i\left(\frac{1}{4c_1}(\mathfrak{f}_{i}(I,I,M)(\LPM{diff}\LPM{rot}+\LPM{rot}\LPM{diff})+\mathfrak{f}_{i}(I,-I,M)\LPM{rot})\right.\nonumber\\
				&\quad+\frac{c_2}{2}\left((\tilde{P}_i\mathfrak{f}_{i}(\mathcal{M}_+^\leftarrow,\mathcal{M}_-^\leftarrow, M)-2\mathfrak{f}_{i}(\mathcal{M}_+^\leftarrow,\mathcal{M}_-^\leftarrow,N))\begin{pmatrix}1&0&0&-1\end{pmatrix}\right.\nonumber\\
				&\quad\left.\left.-\tilde{P}\begin{pmatrix}1\\0\\0\\-1\end{pmatrix}\mathfrak{f}_{i}(\mathcal{M}_+^\rightarrow,\mathcal{M}_-^\rightarrow, M)\right)\right)\bm{w}_i ,
			\end{align}
			and the second-order correction is
			\begin{align}
				\bm{\eta}^{(2)}(\tilde{P},\tilde{t})&=\sum_i\left(\frac{1}{4 c_1^2}\left(\mathfrak{g}_{i,1}(I,I)\ab(\LPM{rot}^2+\frac{\mathcal{M}_{+}+\mathcal{M}_{-}}{2})+\mathfrak{g}_{i,1}(\mathcal{M}_{+},\mathcal{M}_{-})-\mathfrak{h}_{i,1}(M)\right.\right.\nonumber\\
				&\quad+(\tilde{P}-\tilde{P}_i)\mathfrak{h}_{i,0}(M))+\frac{c_2}{2c_1}\ab((\tilde{P}_i\mathfrak{g}_{i,1}(I,I)-\mathfrak{f}_{i}(I,I,M))\LPM{rot}\LPM{mul}+\tilde{P}\mathfrak{g}_{i,1}(I,I)\LPM{mul}\LPM{rot})\nonumber\\
				&\quad+c_2^2\left((\mathcal{M}_{+}^\leftarrow+\mathcal{M}_{-}^\leftarrow)\left(\frac{1}{24}\mathfrak{g}_{i,3}(\mathcal{M}_{+}^\rightarrow,\mathcal{M}_{-}^\rightarrow)-\frac{\tilde{P}}{4}\mathfrak{g}_{i,2}(\mathcal{M}_{+}^\rightarrow,\mathcal{M}_{-}^\rightarrow)\right.\right.\nonumber\\
				&\quad\left.+\ab(\frac{\tilde{P}^2}{2}+\frac{1}{8})\mathfrak{g}_{i,1}(\mathcal{M}_{+}^\rightarrow,\mathcal{M}_{-}^\rightarrow)-\frac{\tilde{P}}{4}\mathfrak{f}_{i}(\mathcal{M}_{+}^\rightarrow,\mathcal{M}_{-}^\rightarrow,M)-\frac{1}{12}\mathfrak{f}_{i}(\mathcal{M}_{+}^\rightarrow,\mathcal{M}_{-}^\rightarrow,N)\right)\nonumber\\
				&\quad+\left(\frac{1}{24}\mathfrak{g}_{i,3}(\mathcal{M}_{+}^\leftarrow,\mathcal{M}_{-}^\leftarrow)+\frac{\tilde{P}_i}{4}\mathfrak{g}_{i,2}(\mathcal{M}_{+}^\leftarrow,\mathcal{M}_{-}^\leftarrow)+\ab(\frac{\tilde{P}_i^2}{2}+\frac{1}{8})\mathfrak{g}_{i,1}(\mathcal{M}_{+}^\leftarrow,\mathcal{M}_{-}^\leftarrow)\right.\nonumber\\
				&\quad\left.-\frac{3\tilde{P}_i}{4}\mathfrak{f}_{i}(\mathcal{M}_{+}^\leftarrow,\mathcal{M}_{-}^\leftarrow,M)+\frac{11}{12}\mathfrak{f}_{i}(\mathcal{M}_{+}^\leftarrow,\mathcal{M}_{-}^\leftarrow,N)\right)(\mathcal{M}_{+}^\rightarrow+\mathcal{M}_{-}^\rightarrow)\nonumber\\
				&\quad+\frac{\tilde{P}}{2}(\mathfrak{f}_{i}(I,I,M)-\tilde{P}_i\mathfrak{g}_{i,1}(I,I))\begin{pmatrix}1\\0\\0\\-1\end{pmatrix}\begin{pmatrix}1&0&0&-1\end{pmatrix}+\frac{1}{2}\mathfrak{h}_{i,2}(N)+\frac{\tilde{P}+3\tilde{P}_i}{2}\mathfrak{h}_{i,1}(N)\nonumber\\
				&\quad+(\tilde{P}_i^2-\tilde{P}^2)\mathfrak{h}_{i,0}(N)-\frac{1}{12}\mathfrak{h}_{i,3}(M)+\frac{\tilde{P}-\tilde{P}_i}{4}\mathfrak{h}_{i,2}(M)-\ab(\frac{\tilde{P}^2+\tilde{P}_i^2}{2}+\frac{1}{4})\mathfrak{h}_{i,1}(M)\nonumber\\
				&\quad\left.\left.+\ab(\frac{\tilde{P}^3-\tilde{P}_i^3}{3}+\frac{\tilde{P}-\tilde{P}_i}{4})\mathfrak{h}_{i,0}(M)\right)\right)\bm{w}_i ,
			\end{align}
			where
			\begin{subequations}
				\begin{align}
					\mathfrak{f}_{i}(\mathcal{T}_+,\mathcal{T}_-,F)&=F_0(\tilde{P}-\tilde{P}_i+c_1\tilde{t})\mathcal{T}_{+}+F_0(\tilde{P}-\tilde{P}_i-c_1\tilde{t})\mathcal{T}_{-}\nonumber\\
					&\quad-F_0(\tilde{P}-\tilde{P}_i)(\mathcal{T}_{+}+\mathcal{T}_{-})\\
					\mathfrak{g}_{i,n}(\mathcal{T}_+,\mathcal{T}_-)&=M_n(\tilde{P}-\tilde{P}_i+c_1\tilde{t})\mathcal{T}_{+}+M_n(\tilde{P}-\tilde{P}_i-c_1\tilde{t})\mathcal{T}_{-}-M_n(\tilde{P}-\tilde{P}_i)(\mathcal{T}_{+}+\mathcal{T}_{-})\nonumber\\
					&\quad+2(N_{n-1}(\tilde{P}-\tilde{P}_i+c_1\tilde{t})\mathcal{T}_{+}+N_{n-1}(\tilde{P}-\tilde{P}_i-c_1\tilde{t})\mathcal{T}_{-}-N_{n-1}(\tilde{P}-\tilde{P}_i)(\mathcal{T}_{+}+\mathcal{T}_{-}))\\
					\mathfrak{h}_{i,n}(F)&=F_n(\tilde{P}-\tilde{P}_i+c_1\tilde{t})\mathcal{M}_{+}+F_n(\tilde{P}-\tilde{P}_i-c_1\tilde{t})\mathcal{M}_{-}
				\end{align}
			\end{subequations}
			for matrix \(\mathcal{T}_\pm\) and function \(F=M,N\). In these equations, the matrices \(\LPM{diff}\), \(\LPM{rot}\), and \(\LPM{mult}\) are defined in Eq.~\ref{eq:mdiff}, Eq.~\ref{eq:mrot}, and Eq.~\ref{eq:mmult}, respectively. The second-order correction is necessary to account for the difference both in the negativity index and in the marginal densities in Fig.~\ref{fig:const neg} and Fig.~\ref{fig:const marg p}.

			In the second case considered in the main text, where the parameters of the constant model are \(D=1\), \(E=0.05\), \(M=2000\), \(P_0=20\) and \(\sigma_P=1\), the strengths of \(\LP{diff}\), \(\LP{mul}\), and \(\LP{rot}\) are \(1\), \(\e{}\), and \(0.2\), respectively. Although only the rotation operation is relatively small, we consider \(\LP{diff}\) to be the subdominant perturbative operator \(\LP{p}\). Taking a Fourier transform
			\begin{equation}
				\bm{\eta}(\xi,\tilde{t})=\frac{1}{\sqrt{2\pi}}\int\bm{\eta}(\tilde{P},\tilde{t})\e{-\iu\tilde{P}\xi}\odif{\tilde{P}} ,
			\end{equation}
			and denoting \(\zeta=\sqrt{c_1^2\xi^2+1}\), the solution is
			\begin{align}
				&\quad\bm{\eta}^{(0)}(\xi,\tilde{t})=\exp(\tilde{t}(\LPM{rot}+\iu c_1\xi\LPM{diff}))\bm{\eta}(\xi,0)\\
				&=\left[I+\frac{\sin(\zeta\tilde{t})}{\zeta}\ab(\iu c_1\xi\LPM{diff}+\LPM{rot})\right.\nonumber\\
				&\quad\left.+\bigg(1-\cos(\zeta\tilde{t})\bigg)\ab(\frac{\iu c_1\xi}{\zeta^2}(\LPM{diff}\LPM{rot}+\LPM{rot}\LPM{diff})+\frac{1}{\zeta^2}(\LPM{diff}^2+\LPM{rot}^2)-\LPM{diff}^2)\right]\bm{\eta}^{(0)}(\xi,0) .
			\end{align}
			As in the case of the exact solution in Eqs.~\ref{eq:const model exact TDSE solution}, \(\bm{\eta}(\tilde{P},\tilde{t})\) may be calculated by numerical Fourier transform.

			For the next order,
			\begin{align}
				&\quad\bm{\eta}^{(1)}(\tilde{P},\tilde{t})=\tint{\e{(\tilde{t}-\tilde{\tau})(\LP{diff}+\LP{rot})}\LP{mul}\bm{\eta}^{(0)}(\tilde{P},\tilde{\tau})}\\
				&=\int\odif{\xi'}\frac{\e{\iu\tilde{P}\xi'}}{\sqrt{2\pi}}\tint{}\e{(\tilde{t}-\tilde{\tau})(\LPM{rot}+\iu c_1\xi'\LPM{diff})}\int\odif{\tilde{P}'}\frac{\e{-\iu\tilde{P}'\xi'}}{\sqrt{2\pi}}2 c_1 c_2\tilde{P}'\LPM{mul}\int\odif{\xi}\frac{\e{\iu\tilde{P}'\xi}}{\sqrt{2\pi}}\bm{\eta}^{(0)}(\xi,\tilde{\tau}) .
			\end{align}
			Since
			\begin{equation}
				\int x\frac{\e{-\iu\xi x}}{\sqrt{2\pi}}\odif{x}=\iu\sqrt{2\pi}\odv{\delta(\xi)}{\xi} ,
			\end{equation}
			by defining \(\xi''=\xi'-\xi\), and integrating by parts, we obtain
			\begin{widetext}
				\begin{align}
					\bm{\eta}^{(1)}(\tilde{P},\tilde{t})&=2c_1 c_2\int\odif{\xi'',\xi}\frac{\e{\iu\tilde{P}(\xi+\xi'')}}{2\pi}\int\odif{\tilde{P}'}\frac{\e{-\iu\tilde{P}'\xi''}}{\sqrt{2\pi}}\tilde{P}'\nonumber\\
					&\quad\times\tint{}\e{(\tilde{t}-\tilde{\tau})(\LPM{rot}+\iu c_1(\xi+\xi'')\LPM{diff})}\LPM{mul}\bm{\eta}^{(0)}(\xi,\tilde{\tau})\\
					&=2c_1 c_2\left[\tilde{P}\int\odif{\xi}\frac{\e{\iu\tilde{P}\xi}}{\sqrt{2\pi}}\left(\frac{\sin(\zeta\tilde{t})}{\zeta}\LPM{mul}+\frac{1-\cos(\zeta\tilde{t})}{\zeta^2}(\LPM{mul}\LPM{rot}+\LPM{rot}\LPM{mul})\right.\right.\nonumber\\
					&\quad\left.+\iu c_1\xi\frac{1-\cos(\zeta\tilde{t})}{\zeta^2}(\LPM{mul}\LPM{diff}+\LPM{diff}\LPM{mul})\right)\bm{\eta}^{(0)}(\xi,0)\nonumber\\
					&\quad+c_1\int\odif{\xi}\frac{\e{\iu\tilde{P}\xi}}{\sqrt{2\pi}}\left(\iu c_1\xi\frac{\sin(\zeta\tilde{t})-\zeta\tilde{t}\cos(\zeta\tilde{t})}{\zeta^3}\LPM{diff}^2\right.\nonumber\\
					&\quad+\frac{\big(1-\cos(\zeta\tilde{t})\big)+(\zeta^2-1)\bigg(\zeta\tilde{t}\sin(\zeta\tilde{t})-(1-\cos(\zeta\tilde{t})\big)\bigg)}{\zeta^4}\LPM{diff}\nonumber\\
					&\left.\left.+\iu c_1\xi\frac{2\big(1-\cos(\zeta\tilde{t})\big)-\zeta\tilde{t}\sin(\zeta\tilde{t})}{\zeta^4}\LPM{rot}\right)\LPM{mul}\bm{\eta}^{(0)}(\xi,0)\right] .
				\end{align}
			\end{widetext}
			The inverse Fourier transform of the equation is done numerically to obtain the results in Fig.~\ref{fig:const as dac neg} and Fig.~\ref{fig:const as dac marg p}.

	\nocite{*}
	\bibliography{qcle-negativity}
	\bibliographystyle{aipnum4-1} 

\end{document}